\documentclass[12pt]{article}
\usepackage{amssymb}
\usepackage{amsfonts}
\usepackage{graphics}
\def\s{\rule{0in}{3.0ex}}
\def\p{\vphantom{\tilde{\phi}}}

\def\beq{\begin{equation}}
\def\eeq{\end{equation}}
\def\beqarray{\begin{eqnarray}}
\def\eeqarray{\end{eqnarray}}
\newenvironment{Eqnarray}{\arraycolsep 0.14em\begin{eqnarray}}{\end{eqnarray}}

\begin{document}

\title{Wavelet Notes}
\author{B. M. Kessler, G. L. Payne, W. N. Polyzou
\\ \it The University of Iowa\\ \it Iowa City, IA, 52242}

\maketitle
\vspace{0.3cm}
\begin{abstract}

Wavelets are a useful basis for constructing solutions of the integral
and differential equations of scattering theory.  Wavelet bases efficiently 
represent functions with smooth structures on different scales, and the 
matrix representation of operators in a wavelet basis are well-approximated 
by sparse matrices.  The basis functions are related to solutions of a 
linear renormalization group equation, and the basis functions have structure 
on all scales.  Numerical methods based on this renormalization group 
equation are discussed.  These methods lead to accurate and efficient 
numerical approximations to the scattering equations.  These notes provide
a detailed introduction to the subject that focuses on numerical methods. 
We plan to provide periodic updates to these notes.  

\end{abstract}
\bigskip
\noindent {\bf file: usr/wavelets/revised-notes.tex}
\bigskip

\section{Introduction}

Wavelets are versatile functions with a wide range of applications
including time-frequency analysis, data compression, and numerical
analysis.  The objective of these notes is to provide an introduction
to the properties of wavelets which are useful for solving integral
and differential equations by using the wavelets to represent the
solution of the equations.

While there are many types of wavelets, we concentrate primarily on
orthogonal wavelets of compact support, with particular emphasis on
the wavelets introduced by Daubechies.  The Daubechies wavelets have
the additional property that finite linear combinations of the
Daubechies wavelets provide local pointwise representations of
low-degree polynomials.  We also have a short discussion 
of continuous wavelets in the Appendix I and spline wavelets in Appendix II.

There notes are not intended to provide a complete discussion of the
subject which can be found in the references given at the end of this
section.  Rather, we concentrate on the specific properties which are
useful for numerical solutions of integral and differential equations.
Our approach is to develop the wavelets as orthonormal basis functions
rather than in terms of low- and high-pass filters, which is more
common for time-frequency analysis applications. 

The Daubechies wavelets have some properties that make them natural 
candidates for basis functions to represent solutions of integral equations.
Like splines, they are functions of compact support that can locally 
pointwise represent low degree polynomials.  Unlike splines, they 
are orthonormal.  More significantly,  only a relatively small
number of wavelets are needed to represent smooth functions.

One of the interesting features of wavelets is that they can be
generated from a single scaling function, which is the solution of a
liner renormalization-group equation, by combinations of translations
and scaling.  This equation, called the scaling equation, expresses
the scaling function on one scale as a finite linear combination of
discrete translations of the same function on a smaller scale.  The
resulting scaling functions and wavelets have a fractal-like
structure.  This means that they have structure on all scales. This
requires a different approach to the numerical analysis, which is
provided by the scaling equation.  These notes make extensive use of
the scaling function.

Some of the references that we have found useful are:

\noindent[1] I. Daubechies, Orthonormal bases of compactly supported wavelets,
Comm. Pure Appl. Math. {\bf 41}(1988)909.

\noindent[2] G. Strang, "Wavelets and Dilation Equations: A Brief
Introduction," SIAM Review, 31:4, pp. 614--627, (Dec 1989).

\noindent[3] I. Daubechies, {\it Ten Lectures on Wavelets},
SIAM, Philadelphia, 1992.

\noindent[4] C. K. Chui {\it Wavelets - A tutorial in Theory and Applications},
Academic Press, 1992. 

\noindent[5] W.-C. Shann, "Quadrature rules needed in
Galerkin-wavelets methods", Proceedings for the 1993 annual meeting of
Chinese Mathematics Association, Chiao-Tung Univ, (Dec 1993).

\noindent[6] W.-C. Shann and J.-C. Yan, "Quadratures involving
polynomials and Daubechies' wavelets", Technical Report 9301,
Department of Mathematics, National Central University, (1993).

\noindent[7] G. Kaiser, {\it A Friendly Guide to Wavelets}, Birkhauser 1994.

\noindent[8] W. Sweldens and R. Piessens, "Quadrature Formulae and
Asymptotic Error Expansions for wavelet approximations of smooth
functions", SIAM J. Numer.  Anal., 31, pp. 1240--1264, (1994).

\noindent[9] H. L. Resnikoff and R. O. Wells, {\it Wavelet Analysis, The Scalable Structure of Information}, Springer Verlag, NY.

\noindent[10]  O. Bratelli and P. Jorgensen, 
{\it Wavelets through a Looking Glass},
Birkhauser, 2002.

\bigskip
In addition, some of the material in these notes is in our paper
\bigskip

\noindent[11] B. M. Kessler, G. L. Payne, and W. N. Polyzou, Scattering 
Calculations With Wavelets, Few Body Systems, 33,1-26(2003). 

\section{Haar Scaling Functions and Wavelets}

Scaling functions play a central role in the construction of
orthonormal bases of compactly supported wavelets.  The scaling 
functions and wavelets are distinct bases related by an orthogonal
transformation called the wavelet transform.  

The concept of scaling functions is most easily understood using 
Haar wavelets (these are made out of simple box functions).  The Haar 
functions are the simplest compactly supported scaling functions 
and wavelets. 

The {\bf Haar scaling function} is defined by
\beq
\phi (x) := \left\{
\begin{array}{ll}
0 & x \leq 0 \\
1 & 0<x\leq 1 \\
0 & x>1 
\end{array} 
\right. \,.
\eeq

It satisfies the normalization conditions: 
\beq
(\phi,\phi) := \int^{\infty}_{-\infty} \phi^* (x) \phi (x) dx =
\int_0^1 \phi (x) dx =1.
\eeq

The operations of discrete translation and dilatation are used 
extensively in the study of compactly supported wavelets.  
The {\bf unit translation operator $T$} is defined by 
\beq 
(T \chi )(x) = \chi (x-1).  
\eeq 
This operator translates the function $\chi (x)$ to the right by one unit.
The unit translation operator has the property: 
\beq 
( T \psi , T \chi) = \int^{\infty}_{-\infty} \psi^* (x-1) \chi (x-1) dx= 
\eeq
\beq 
\int^{\infty}_{-\infty} \psi^* (y) \chi (y) dy= (\psi, \chi) 
\eeq
where $y=x-1$.  This means that the unit translation operator preserves 
the scalar product:
\beq
( T \psi , T \chi) = ( \psi ,  \chi). 
\eeq

If $A$ is a linear operator its {\bf adjoint} $A^{\dagger}$ is defined by 
the relation
\beq
(\psi , A^{\dagger} \chi ) = ( A \psi , \chi ). 
\eeq
It follows that 
\beq
( \psi , T^{\dagger} \chi ) = (T\psi, \chi ) =
\int^{\infty}_{-\infty} \psi^* (x-1) \chi (x) dx .
\eeq
Changing variables to $y=x-1$ gives
\beq
( \psi , T^{\dagger} \chi ) = 
\int^{\infty}_{-\infty} \psi^* (y) \chi (y+1) dy 
\eeq
or
\beq
(T^{\dagger} \chi ) (x) = \chi (x+1) 
\eeq
which is a left shift by one unit.
Since
\beq
(\psi , \chi ) = (T\psi , T\chi )= (\psi ,T^{\dagger} T \chi )
\eeq
it follows that $T^{\dagger}= T^{-1}$.  An operator whose adjoint is its
inverse is called unitary.  Unitary operators preserve inner products.

It follows from the definition of the Haar scaling function, $\phi (x)$, that 
\[
(T^m \phi , T^n \phi ) = ( \phi , T^{n-m}\phi)=
\int^{\infty}_{-\infty} \phi^* (x) \phi (x-n+m) dx =
\]
\beq
\int^{1}_{0} \phi (x-n+m) dx =
\delta_{nm}    
\eeq
This means the functions 
\beq
\phi_n (x) := (T^n \phi)(x) = \phi (x-n) 
\eeq
are orthonormal.  There are an infinite number of these functions 
for integers $n$ satisfying $-\infty < n < \infty$.

The integer translates of the scaling function span a space,
${\cal V}_0$, which is a subspace of the space of square integrable functions.
The elements of ${\cal V}_0$ are functions of the form
\beq
f(x) = \sum_{n=-\infty}^{\infty} f_n \phi_n (x) =
\sum_{n=-\infty}^{\infty} f_n (T^n\phi) (x)
= \sum_{n=-\infty}^{\infty} f_n \phi (x-n), 
\eeq
where the square integrability requires that the coefficients satisfy 
\beq
\sum_{n=-\infty}^{\infty} \vert f_n \vert^2 < \infty.
\eeq
For the Haar scaling function ${\cal V}_0$ is the space of square
integrable functions that are piecewise constant on each unit-width
interval.  Note that while there are an infinite number of functions
in ${\cal V}_0$, it is a small subspace of the space of square
integrable functions.

In addition to translations $T$,  the linear operator $D$, 
corresponding to {\bf discrete scale transformations}, is defined by:
\beq
(D \chi )(x) = {1 \over \sqrt{2}} \chi (x/2).
\eeq
When this is applied to the Haar scaling function it gives 

\beq
(D\phi) (x) = \left\{
\begin{array}{ll}
0 & x \leq 0 \\
{1 \over \sqrt{2}} & 0<x\leq 2 \\
0 & x>2 
\end{array} 
\right. .
\eeq

This function has the same box structure as the original Haar scaling 
function, except it is 
twice as wide as the original scaling function and shorter 
by a factor of $\sqrt{2}$.  Note that the normalization ensures 
\beq
(D\psi , D\chi ) = 
\int^{\infty}_{-\infty} {1 \over 2}\psi^* (x/2) \chi (x/2) dx 
\eeq
\beq
= \int^{\infty}_{-\infty} \psi^* (y) \chi (y) dy =
(\psi , \chi ) 
\eeq
where the variable in the integrand has been changed to $y=x/2$.

The adjoint of $D$ is determined by the definition 
\beq
(\psi , D^{\dagger} \chi ) = (D\psi , \chi ) =
\int^{\infty}_{-\infty} {1 \over \sqrt{2}}\psi^* (x/2) \chi (x) dx.
\eeq
Setting $y=x/2$ gives
\beq
\int^{\infty}_{-\infty}\psi^* (y)\sqrt{2} \chi (2y) dy 
\eeq
which gives 
\beq
(D^{\dagger}\chi )(x)= \sqrt{2} \chi (2x). 
\eeq
This shows that $D^{\dagger} = D^{-1}$ or $D$ is also unitary.

Define the functions constructed by $n$ translations followed by
$m$ scale transformations 
\beq
\phi_{mn} (x) = (D^m T^n \phi) (x) = (D^m \phi_n) (x)
\eeq
\beq
= 2^{-m/2} \phi (2^{-m}x -n) = 2^{-m/2} \phi (2^{-m}(x - 2^m n)).  
\eeq
It follows that for a fixed scale $m$ 
\beq
(\phi_{mn} , \phi_{mk} ) = (D^m \phi_n , D^m \phi_k ) = (\phi_n ,
D^{m-m} \phi_k)= (\phi_n , \phi_k) = \delta_{nk}.
\eeq
This shows that the functions $\phi_{mn}(x)$ for any  fixed scale 
$m$ are orthonormal.  

We define the subspace ${\cal V}_m$ of the square integrable functions 
to be those functions of the form:
\beq
f(x) = \sum_{n=-\infty}^{\infty} f_n \phi_{mn} (x) =
\sum_{n=-\infty}^{\infty} f_n (D^mT^n\phi) (x) 
\eeq
where the square integrability requires that the coefficients satisfy 
\beq
\sum_{n=-\infty}^{\infty} \vert f_n \vert^2 < \infty.
\eeq
These elements of ${\cal V}_m$ are square summable functions 
that are piecewise constant
on intervals of width $2^m$.  The spaces ${\cal V}_m$ and 
${\cal V}_0$ are related by $m$ scale transformations
$D^m {\cal V}_0={\cal V}_m$. 

In general the scaling function $\phi (x)$ is defined as the solution
of a scaling equation subject to a normalization condition.  The
scaling equation relates the scaled scaling function, $(D\phi) (x)$,
to translates of the original scaling function.  The general form of
the {\bf scaling equation} is
\beq
(D\phi) (x) = \sum_l h_l T^l \phi (x)
\label{eq:AAxx}
\eeq
where $h_l$ are fixed constants, and the sum may be finite or infinite.
This equation can be expressed as
\beq
{1 \over \sqrt{2} } \phi ({x\over 2}) = \sum_l h_l \phi (x-l)
\eeq
which is sometimes written as
\beq
\phi (x) = \sqrt{2} \sum_l  h_l \phi (2 x-l) =
\sum_l  c_l \phi (2 x-l)
\eeq  
where $c_l = \sqrt{2} h_l$.  Equation (\ref{eq:AAxx}) is the most important 
equation in these notes.

In general the scaling equation cannot be solved analytically.  In 
the special case of the Haar scaling function 
the solution is obtained by observing that the scaled box is
stretched over two adjacent boxes with a suitable reduction in height.
It follows that:
\[
D\phi (x) = {1 \over \sqrt{2}} \phi (x/2) =
{1 \over \sqrt{2}} \phi (x) + {1 \over \sqrt{2}}T \phi (x) 
\]
\beq
={1 \over \sqrt{2}} \phi (x) + {1 \over \sqrt{2}}\phi (x-1) .
\eeq
Here $h_0 = h_1 = 1/\sqrt{2}$.  These coefficients are special to the
Haar scaling function.  The best way to think about the scaling
function $\phi (x)$ is to note that the scaling function $\phi (x)$ is
the solution of the scaling equation up to normalization.  The
normalization is fixed by
\[
\int \phi (x) dx =1 .
\]

An additional relation involving the translation  
$T$ and dilatation operator $D$ is useful for future 
computations.  First note that

\beq
DT \psi (x) = D \psi (x-1) = {1 \over \sqrt{2} } \psi (x/2-1) =
 {1 \over \sqrt{2} } \psi ({x-2 \over 2} ) =
T^2 D \psi (x), 
\eeq
which leads to the operator relation 
\beq
DT = T^2 D . 
\eeq
It follows from this equation that 
\beq
D\phi_n(x)  = D T^n \phi (x)= T^{2n}D \phi (x) =
T^{2n} (h_0 \phi (x) + h_1 T \phi (x) ). 
\eeq
This shows that all of the basis elements in ${\cal V}_1$ can be
expressed in terms of basis elements in ${\cal V}_0$.  For the case of
the Haar scaling function this is obvious, but the argument above is
more general.

Specifically if $\psi(x)  \in {\cal V}_1$ then
\beq 
\psi (x) = \sum_{n=-\infty}^{\infty} d_n \phi_{1n}(x) 
= \sum_{n= -\infty}^{\infty} d_n D \phi_{n}(x)
\eeq
\beq
= \sum_{n= -\infty}^{\infty} [d_n h_0 \phi_{2n}(x) + d_n h_1 \phi_{2n+1}(x)]
= \sum_{-\infty}^{\infty}  e_n \phi_{n}(x) 
\eeq 
where 
\beq
e_{2n} = d_n h_0 \qquad e_{2n+1} = d_n h_1 .
\eeq
It is easy to show that 
\beq
\sum_{n=-\infty}^{\infty}  \vert e_n \vert^2 =
\sum_{n=-\infty}^{\infty}  \vert d_n \vert^2 .
\eeq

What we have shown, as a consequence of the scaling equation,  is 
the inclusion property 
\beq
{\cal V}_0 \supset {\cal V}_1 .
\eeq
Similarly, using the same method, it is possible to show the 
chain of inclusions 
\beq
\cdots {\cal V}_{-k} \supset {\cal V}_{-k+1} \supset \cdots \supset {\cal V}_0 
\supset \cdots {\cal V}_k \supset {\cal V}_{k+1} \cdots
\eeq
These properties hold for the solution of any scaling equation. 
In the Haar example the spaces ${\cal V}_m$ are spaces of piecewise 
constant, square integrable functions that are constant on intervals 
of the real line of width $2^m$.
  
The subspaces ${\cal V}_m$ are used as approximation spaces in 
applications.   To understand how they are used as approximation 
spaces note that as $m \to -\infty$ the approximation to $f(x)$ given by 
\beq
f_m (x) = \sum_{n=-\infty}^{\infty} f_{mn} \phi_{mn}(x)  
\eeq
with 
\beq
f_{mn} = \int_{-\infty}^{\infty} \phi_{mn}(x) f(x) dx
\eeq
is bounded by the upper and lower Riemann sums for steps of width
$2^{-m}$. This is because, up to a scale factor,  the coefficients 
$f_{mn}$ are just average values of the function on the appropriate 
sub-interval (to deal with the infinite interval it is best to first 
consider functions that vanish outside of finite intervals and take 
limits).  Since the upper
and lower Riemann sums converge to the same integral (when the
function is integrable) it follows that

\beq
\int_{-\infty}^{\infty} \vert f_m(x)-f(x)\vert dx < \epsilon
\eeq
for sufficiently large $-m$.  A similar argument can be extended
to get $L^2$ convergence .

Similarly, as $m \to + \infty$, the width of $\phi_{mn}(x)$ grows 
like $2^m$ while the height falls off like $2^{-m/2}$.  Again,
if the function vanishes outside of a bounded interval then for 
sufficiently large $m$ there is only one (or two) 
$\phi_{mn}(x)$ that are non-vanishing where the function is non-vanishing.
In the case that only one $\phi_{mn}=\phi_{mn_0}$ overlaps the support 
of $f(x)$
\beq
f_m (x) \sim 2^{-m/2} \phi_{mn_0} (x) \int_{-\infty}^{\infty} f(x) dx . 
\eeq
The integral of the square of this function $\sim 2^{-m} \to 0$ as
$m \to \infty$.

Note that 
\beq
\int_{-\infty}^{\infty} f_m(x) dx \to \int_{-\infty}^{\infty} f(x)dx
\eeq
as $m \to \infty$.  This shows that the limit of the integral of
$f_m(x)$ as $m\to \infty$  
is finite in $L^1$ but
$0$ in $L^2$. 

It is useful to express some of these results in a more useful form.
Define the projection operators
\beq
P_m f(x) = \sum_{n=-\infty}^{\infty} f_{mn} \phi_{mn} (x) 
\eeq
where
\beq
f_{mn} = \int_{- \infty}^{ \infty}\phi_{mn}^* (x) f(x) dx .
\eeq
The above conditions can be stated in terms of these projectors:
\beq
\lim_{m \to - \infty } P_m  = I     
\label{eq:comp}
\eeq
\beq
\lim_{m \to + \infty } P_m  = 0 .     
\label{eq:null}
\eeq
These results mean that the approximation space ${\cal V}_m$
approaches the space of square integrable functions as $m\to -\infty$.
We have shown that (\ref{eq:comp}) and (\ref{eq:null}) are valid for
the Haar scaling function, but they are also valid for a large class of scaling
functions,

We are now ready to construct {\bf wavelets}.  First recall the condition
\beq
{\cal V}_0 \supset {\cal V}_1 .
\eeq
Let ${\cal W}_1$ be the subspace of vectors in the space ${\cal V}_0$ that 
are orthogonal to the vectors in ${\cal V}_1$.  We can write
\beq
{\cal V}_0 = {\cal V}_1 \oplus {\cal W}_1.
\eeq
This notation means that any vector in ${\cal V}_0$ can be expressed as
a sum of two vectors - one that is in ${\cal V}_1$ and one that is 
orthogonal to every vector in ${\cal V}_1$.

Note that the scaling equation implies that 
every vector in ${\cal V}_1$ can be expressed as a linear 
combination of vectors in ${\cal V}_0$ using
\beq
D\phi_n (x) = h_0 \phi_{2n} (x) + h_1 \phi_{2n+1} (x). 
\eeq
Clearly the functions that are orthogonal to these in ${\cal V}_1$
on the same interval can be expressed in terms of the 
difference functions
\beq
\psi_{1n} (x) := D\psi_{n} (x) = h_1 \phi_{2n} (x) - h_0 \phi_{2n+1} (x) = 
{1 \over \sqrt{2}} ( \phi_{2n} (x) -  \phi_{2n+1} (x)) .
\eeq
Direct computation shows that the $\psi_{1n}(x)$  are elements 
of ${\cal V}_0$ that satisfy
\beq
(D\psi_{1n} , D\phi_l )  =0 .
\eeq
and 
\beq
(\psi_{1n} , \psi_{1k} ) = \delta_{nk} .
\eeq

Thus we conclude that ${\cal W}_1$ is the space of square integrable 
functions of the form
\beq
f(x )= \sum_{n=-\infty}^{\infty} f_n \psi_{1n} (x)  
\eeq
with 
\beq
f(x )= \sum_{n=-\infty}^{\infty} \vert f_n \vert^2 . 
\eeq

Similarly, we can decompose ${\cal V}_l = {\cal V}_{l+1} \oplus {\cal W}_{l+1}$
for each value of $l$.  For the special case of ${\cal W}_0$ 
we define the Haar {\bf mother
wavelet} as
\beq
\psi(x) := D^{-1} ( h_1 \phi (x) - h_0 T \phi (x) )
= 
\eeq
\beq
h_1 \sqrt{2} \phi (2t) - h_0 \sqrt{2} \phi (2(t-1))=
( \phi (2t) - \phi (2(t-1)))
\eeq
which is manifestly orthogonal to the scaling function.  Translates of 
the mother wavelet define a basis for ${\cal W}_0$
\beq
\psi_n (x) = T^n \psi(x) =  T^n D^{-1} ( h_1 \phi (x) - h_0 T \phi (x) )=
\eeq
\beq
D^{-1} ( h_1 \phi_{2n} (x) - h_0  \phi_{2n+1} (x) ).
\eeq

If we decompose ${\cal V}_m$ we have:
\[
{\cal V}_{-m} =  {\cal W}_{-m+1} \oplus {\cal V}_{-m+1}  
\]
\[
={\cal W}_{-m+1} \oplus {\cal W}_{-m+2} \oplus {\cal V}_{-m+2} 
\]
\beq
={\cal W}_{-m+1} \oplus {\cal W}_{-m+2} \oplus \cdots \oplus {\cal W}_l 
\oplus {\cal V}_{l}. 
\label{eq:mres}
\eeq
Note that unlike the ${\cal V}_m$ spaces, the ${\cal W}_m$ spaces are
all mutually orthogonal, since if $m>n \to {\cal W}_m \subset {\cal V}_n$ 
which is orthogonal to ${\cal W}_n$ by definition.

If $f(x)$ is any square integrable function the conditions
\beq
\lim_{m \to - \infty } P_m  = I     
\label{eq:lima}
\eeq
\beq
\lim_{m \to + \infty } P_m  = 0     
\label{eq:limb}
\eeq
mean that for sufficiently large $m$ and any $l$ that $f(x)$ can be well 
approximated by a function in 
\beq
{\cal W}_{-m+1} \oplus {\cal W}_{-m+2} \oplus \cdots \oplus {\cal W}_l .
\eeq
This means that the function can be approximated by a linear combination of 
basis functions (wavelets) from each of the spaces ${\cal W}_r$ .

A {\bf multiresolution analysis} is a set of subspaces ${\cal V}_m$ and 
${\cal W}_m$ satisfying (\ref{eq:mres}), (\ref{eq:lima}), 
and (\ref{eq:limb}).  The condition (\ref{eq:lima}) allows one to 
interpret the space ${\cal V}_{m}$, for sufficiently large $-m$, as 
an approximation space for numerical applications.

Basis functions for ${\cal W}_m$ are given by
\beq
\psi_{mn}(x) = D^m T^n \psi(x) = 
D^{m-1} ( h_1 \phi_{2n} (x) - h_0  \phi_{2n+1} (x) ) .
\label{eq:mom}
\eeq
That these are a basis with the required properties is easily shown by 
showing that these functions are orthogonal to ${\cal V}_{m}$ and 
can be used to recover the remaining vectors in ${\cal V}_{m-1}$. 

The functions $\psi_{nl} (x) $, are called Haar wavelets. They satisfy
the orthonormality conditions:
\beq
(\psi_{nl} , \psi_{n'l'} ) = \delta_{nn'} \delta_{ll'} 
\eeq
where the $\delta_{nn'}$ follows from the orthogonality of the 
spaces ${\cal W}_n$ and ${\cal W}_{n'}$ for $n\not=n'$.

The $\delta_{ll'}$ follows from the unitarity of $D$ and 
\beq
(\psi , T^n \psi ) =\delta_{n0}.
\eeq

The important steps discussed above generalize to the case of a 
general scaling equation of the form:
\beq
D\phi (x) = \sum h_l T^l \phi (x).
\eeq
This equation is
solved to find the scaling function $\phi (x)$.  This, along with translations
and dilatations is used to construct the spaces ${\cal V}_l$.   The 
scaling equation ensures the existence of spaces ${\cal W}_m$,
satisfying ${\cal V}_{m+1}={\cal W}_{m}\oplus {\cal V}_{m}$  that can
be used to build discrete orthonormal bases.  The mother wavelet 
function is expressed in terms of the scaling function and 
the coefficients $h_l$
as 
\beq
\psi(x) = D^{-1} \sum_l g_l T^l \phi (x)
\eeq
where we will see later that
\beq
\qquad g_l = (-)^kh_{k-l}  
\eeq
where $k$ is any odd integer.  In general the coefficients $h_l$ must
satisfy constraints for the solution to the scaling equation to exist.
General wavelets can be expressed in terms of the mother wavelet using
(\ref{eq:mom}).  In the next section the coefficients $g_l$ will be
expressed in terms of the scaling function.

\section{Scaling Functions - General Considerations}

This section extends the treatment of scaling equation to a 
more general class of scaling functions than the Haar functions.
In general, a scaling function satisfies the following three
conditions.  First, 
the {\bf scaling function} is the solution of the {\bf 
scaling equation}
\beq
D \phi (x) = \sum_l h_l T^l \phi (x)
\label{eq:iaa}
\eeq
where $h_l$ are numerical coefficients that define the scaling equation.
Second, in addition to satisfying the scaling equations, 
{\bf integer translates} of the scaling functions are required to be 
{\bf orthonormal}
\beq
(\phi_n , \phi_m ) = (T^n \phi , T^m \phi ) = ( \phi , T^{m-n}\phi )
= \delta_{mn}. 
\label{eq:norma}
\eeq
Third, the initial scale is fixed by the {\bf normalization condition}
\beq
\int \phi (x) dx =1.
\label{eq:normb}
\eeq
It might seem like the normalization conditions in (\ref{eq:norma})
and (\ref{eq:normb}) are not compatible.  To see that this is not true  
note that condition (\ref{eq:norma}) is invariant under unitary 
changes of scale of the form
\[
D_s \chi (x) := {1 \over \sqrt{s}} \phi \left ({x \over s} \right )
\]
while condition (\ref{eq:normb}) is not.  It follows that
condition (\ref{eq:normb}) can be interpreted as setting a starting 
scale, $s$.  The condition (\ref{eq:norma}) is preserved independent 
of the starting scale.

We now investigate the consequences of these three conditions.
Using the definitions of the operators $D$ and $T$ the scaling
equation becomes:
\beq
{1 \over \sqrt{2}} \phi ({x \over 2}) = \sum h_l \phi (x-l).
\label{eq:iab}
\eeq
As shown in section 1, it can be put in the useful form
\beq
\phi (x) = \sum_l \sqrt{2} h_l \phi (2x-l).
\label{eq:iac}
\eeq
In general the sums may be from $-\infty \to \infty$.  Finite sums are treated
by assuming that only a finite number of the $h_l$'s are non zero.  All of the 
compactly supported scaling functions are solutions of scaling equations with 
a finite number of non-zero coefficients.

If the scaling equation has a solution, it is unique up to
an overall normalization factor.  To see this take the
Fourier transform of both sides of equation (\ref{eq:iac}) to get
\beq
\tilde{\phi}(k)= {1 \over \sqrt{2 \pi} }
\int_{-\infty}^{\infty}  e^{-i k x} \phi (x) dx = \sum_l \sqrt{2} h_l 
{1 \over \sqrt{2 \pi} }
\int_{-\infty}^{\infty}  e^{-i k x} \phi (2x-l) dx.
\label{eq:iad}
\eeq

Changing variables $x \to 2x-l$ on the right-hand side gives
\beq
{1 \over \sqrt{2 \pi} }
\int_{-\infty}^{\infty}  e^{-i k x} \phi (x) dx = \sum_l {1 \over \sqrt{2}} h_l 
{1 \over \sqrt{2 \pi} }
\int_{-\infty}^{\infty}  e^{-i (k/2) (x+l) } \phi (x) dx
\label{eq:iae}
\eeq
or 
\beq
\tilde{\phi} (k) = \tilde{\phi}\left ({k \over 2}\right )  
\tilde{h}\left ( {k \over 2}\right )  
\label{eq:iaf}
\eeq
where  
\beq
\tilde{h}({k})  =
\sum_l {h_l \over \sqrt{2}} e^{-i k l }.
\label{eq:iag}
\eeq

This form of the scaling equation can be iterated $n$ times to get:
\beq
\tilde{\phi} (k) = \tilde{\phi}\left ({k \over 2^n}\right ) \prod_{m=1}^n 
\tilde{h}\left ({k \over 2^m}\right ) 
\label{eq:iah}
\eeq

This equation holds for any $n$ provided the Fourier transforms exist.
For a finite $n$, an approximation can be made by a finite number of
iterations of the form
\beq
\tilde{\phi}_n (k) = \tilde{\phi}_{n-1}\left ({k \over 2}\right )  
\tilde{h}\left ({k \over 2}\right )  
\label{eq:iai}
\eeq
for any starting function $\tilde{\phi}_0 (k)$.  In the limit of large $n$ 
the function $\tilde{\phi}_n (k)$ should converge to a solution to 
the scaling equation.
The result of formally taking this limit is 
\[
\tilde{\phi} (k) = \lim_{n\to \infty} \tilde{\phi}_0({k \over 2^n}) 
\prod_{l=1}^n  \tilde{h}({k \over 2^l}) 
\]
\beq
=\tilde{\phi}(0) 
\prod_{l=1}^{\infty}  \tilde{h}({k \over 2^l}) .
\label{eq:iaj}
\eeq
If the limit exists as $n \to \infty$,  and the scaling function 
is continuous in a neighborhood of zero,  then the solution 
of the scaling equation is 
uniquely determined by the scaling coefficients $h_l$ up to the 
overall normalization $\tilde{\phi}_0(0)$.  The 
condition $\tilde{\phi}_0(0)=1/\sqrt{2\pi}$ 
is equivalent to the standard normalization condition 
\[
\int_{-\infty}^{\infty} \phi (x) dx=1.
\]
The resulting solution of the scaling equation is independent of the choice 
of starting function provided it is normalized so 
$\tilde{\phi(0)} =1/\sqrt{2\pi}$.  
Once the normalization is fixed, the limit only depends on the 
coefficients $h_l$.  

Thus, if the infinite product converges, then we have an expression for the 
scaling function, up to normalization, which is fixed by assigning a value
to $\tilde{\phi} (0)$.  To show how this works we compute this limit for the
Haar scaling equation.

For the Haar scaling equation the expression for the 
scaling function is 
\[
{1 \over \sqrt{2 \pi}}  
\prod_{l=1}^{\infty}  {1 \over 2} (1 + e^{-ik/2^l}) 
\]
\[
=\lim_{l\to \infty} 
{1 \over \sqrt{2 \pi}}   {1 \over 2^l} (1+ e^{-ik/2})(1+ e^{-ik/4})\cdots
(1+ e^{-ik/2^l})
\]
expanding this out in powers of $e^{-ik/2^l}$ gives
\[
={1 \over \sqrt{2 \pi}}   \lim_{l \to \infty} 
{1 \over 2^l} \sum_{m=0}^{2^{l}-1} (e^{-ik/2^l})^m
\]
\[
={1 \over \sqrt{2 \pi}}   \lim_{l \to \infty} {1 \over 2^l}
{1-e^{-ik} \over  1-e^{-ik/2^l}} =
\]
\beq
={1 \over \sqrt{2 \pi}}   e^{-ik/2}{\sin (k/2) \over (k/2)} .
\label{eq:iaja}
\eeq
A direct calculation of the Fourier transform of the Haar scaling function 
gives
\[
\tilde{\phi}(k) = {1 \over \sqrt{2\pi}}\int_{-\infty}^\infty e^{-ikx} \phi (x)dx= {1 \over \sqrt{2\pi}}\int_0^1 e^{-ikx} dx
\]
\beq
={1 \over \sqrt{2\pi}}
e^{-ik/2}{\sin (k/2) \over (k/2)}
\eeq
which agrees with (\ref{eq:iaja}).

The above analysis shows that the solution of the scaling equation 
depends on the choice of scaling coefficients $h_l$.
The scaling coefficients $h_l$ are not arbitrary.  First note that setting 
$k=0$ in (\ref{eq:iaj}) gives  
\beq
1 = \prod_{l=0}^\infty  \tilde{h}(0). 
\label{eq:iak}
\eeq  
Now using (\ref{eq:iag}) gives 
\beq
\tilde{h} (0) = 1 =  \sum_l {h_l \over \sqrt{2}}
\eeq
or
\beq
\sum_l h_l =  \sqrt{2}.
\label{eq:ial}
\eeq
This condition is satisfied by the Haar wavelets.   This is a 
necessary condition on the scaling coefficients in order to have a 
solution to the scaling equation.

Another condition which constrains the scaling coefficients   
is the orthogonality of the unit translates, 
$(\phi_n, \phi_m) = \delta_{nm}$.  This requires, using (\ref{eq:iac}),
\[
2 \sum_{lk} h_l h_k \int_{-\infty}^{\infty} \phi (2x-2n-l)\phi(2x-2m-k)dx
\]
\[
=2 \sum_{lk} h_l h_k \int_{-\infty}^{\infty} \phi (2x)\phi(2x-2(m-n)-(k-l))dx 
\]
\[
=\sum_{lk} h_l h_k \int_{-\infty}^{\infty} \phi (x)\phi(x-2(m-n)-(k-l))dx
\]
\beq
=\sum_l h_l h_{l-2(m-n)} = \delta_{mn} 
\label{eq:iam}
\eeq
or equivalently 
\beq
\sum_l h_{l-2m} h_{l} = \delta_{m0} .
\label{eq:ian}
\eeq
This is trivially satisfied for the Haar wavelets.  Here and in
all that follows we restrict our considerations to the case that the 
scaling coefficients and scaling functions are real. 

The orthogonality condition also requires that the number of 
non-scaling coefficients must be even.  To see this assume by 
contradiction that there are $2N+1$ non-zero scaling 
coefficients,  $h_0 \cdots h_{2N}$. Then setting 
$m=-N$ in (\ref{eq:ian}) gives 
\beq
\sum_l h_{l+2N} h_{l} = h_{2N} h_{0} = \delta_{N0} =0 .
\eeq
which means that either $h_0=0$ or $h_{2N}=0$, which contradicts 
the assumption that there are $2N+1$ non-zero scaling coefficients.  
This shows that if the number of non-zero scaling coefficients are finite, 
then there must be an {\it even number}, $2K$, with $l=0\cdots 2K-1$.  

Note that if there are only two non-vanishing scaling coefficients,  
$h_0$ and $h_1$,
then the conditions (\ref{eq:ial}) and (\ref{eq:ian}) have a unique 
solution, which is the Haar scaling coefficients.  In this 
case these equations become
\beq
h_0 + h_1= \sqrt{2} 
\eeq
\beq
h_0 h_0 + h_1 h_1 = 1 .
\eeq
These equations have the unique solution $h_0 = h_1 = 1 / \sqrt{2}$.

Conditions (\ref{eq:ial})
and (\ref{eq:ian}) are important constraints on the scaling 
coefficients.

For scaling equations with more than two non-zero scaling
coefficients, additional conditions are needed to determine the
scaling coefficients.

The number of non-zero scaling coefficients determines the support of
the scaling function.  The important property is that scaling
functions that are solutions of a scaling equation with a finite
number of non-zero scaling coefficients have compact support.  The
support is determined by the number of non-zero scaling coefficients.

To determine the support of the scaling function, consider a 
scaling equation with $N=2K+1$ non-zero scaling coefficients.
The scaling function is given by
\[
\phi(x) = {\tilde{\phi}(0)  \over \sqrt{2 \pi}} \int_{-\infty}^{\infty} 
e^{i k x} \prod_{m=1}^{\infty}  \tilde{h}({k \over 2^m}) dk 
\]
\[
={1 \over 2 \pi} \int_{-\infty}^{\infty} 
e^{i k x} \prod_{m=1}^{\infty} \left (\sum_{n_m=0}^{N-1} {h_{n_m} \over \sqrt{2}} 
e^{-i k {n_m}/2^m }\right ) dk
\]
\[
=\lim_{m\to \infty}  
\sum_{n_1=0}^{N-1} \cdots \sum_{n_m=0}^{N-1} \left ( \prod_{k=1}^m 
{h_{n_k} \over \sqrt{2}}\right )
\delta (x -\sum_{k=1}^m  n_k/2^m ).
\]

This defines the scaling function as a distribution.  
This is not a useful representation for computation, however it indicates
that if a scaling function has $N$ non-zero coefficients $h_l$ then the 
scaling function has support on 
\[
[0,(N-1)({1 \over 2} + {1 \over 4} + {1 \over 8} \cdots )] = [0,N-1]
\]
where $N$ is the number of non-zero scaling coefficients.

While the support condition depends only on the number of non-zero 
coefficients, there are many scaling functions with $N$ 
non-zero scaling coefficients.   Except for the constraints dictated by 
the scaling equation,  orthonormality, and normalization, there is 
considerable freedom in choosing the coefficients $h_l$. 

The scaling coefficients also determine the mother wavelet function.  In the general
case the spaces ${\cal V}_m$ are the spaces of square integrable
functions spanned by the orthogonal basis functions $\phi_{mn}(x) :=
D^m T^n \phi (x)$ for integer $n$ satisfying $-\infty < n < \infty$.
As in the Haar case, the scaling equation implies that ${\cal V}_m
\supset {\cal V}_{m+k}$ for $k>0$.  Wavelet spaces are defined by
\[
{\cal W}_m : {\cal V}_{m-1} = {\cal V}_{m}\oplus {\cal W}_{m} . 
\]
It they also satisfy (\ref{eq:lima}) and (\ref{eq:limb}) they 
define a {\bf multiresolution analysis}.
The 
{\bf mother 
wavelet function} lives in the space ${\cal W}_0$ which means that it has an 
expansion in ${\cal V}_{-1}$:
\beq
\psi (x)  = \sum_{n} \sqrt{2} g_n \phi (2x-n )=
\sum_{n}  g_n D^{-1} T^n \phi (x).
\label{eq:iau}
\eeq
This equation can be expressed in a form similar to the scaling equation:
\beq
D \psi (x) =\sum_{n}  g_n  T^n \phi (x).
\label{eq:iav}
\eeq
  
The mother wavelet and all of its integer translates should be orthogonal 
to the scaling function, which is in ${\cal V}_0$.
In terms of the coefficients this requirements is: 
\[
(\psi_m , \phi)   = \sum_{n,l} h_l g_n (\phi_{n+2m},\phi_l ) 
\]
\[
=\sum_{n,l} h_l g_n \delta_{n+2m,l} 
\]
\beq
=\sum_{n} h_{n+2m} g_n  =0 
\label{eq:iava}
\eeq
for all $m$.  

Orthonormality of the translated mother function requires 
\[
(\psi_m , \psi_n)   = \sum_{l,k} g_l g_k (\phi_{l+2m},\phi_{k+2n} ) 
\]
\beq
\sum_{k} g_{k+2(n-m)} g_k = \delta_{mn} 
\label{eq:iaw}
\eeq
or equivalently
\beq
(\psi_m , \psi)   =
\sum_{k}  g_{k+2m} g_k = \delta_{m0}. 
\label{eq:iax}
\eeq
The choice
$g_k := (-1)^k h_{l-k}$ where $l$ is any odd integer it 
satisfies (\ref{eq:iav}) and 
(\ref{eq:iax}):
\[
\sum_{k}  g_{k+2(n-m)} g_k = 
\sum_{k}   (-1)^{k+2(n-m)} h_{l-k-2(n-m)} (-1)^k h_{l-k} 
\]
\beq 
= \sum_{k'}   h_{k'+2(n-m)} h_{k'} = 
\delta_{mn} 
\label{eq:iay}
\eeq
where we have let $k'= l-k$ in the last term.  It also follows that 
\[
\sum_{n} h_{n+2m} g_n  =
\sum_{n} h_{n+2m} (-1)^n h_{l-n}  
\]
\[
=\sum_{n'} h_{l-n'} (-1)^{l-n'-2m} h_{n'+2m}  = 
(-)^l\sum_{n'} h_{l-n'} (-1)^{n'} h_{n'+2m}  
\]
\beq
= (-1)^l\sum_{n'} g_{n'} h_{n'+2m} .   
\label{eq:iaz}
\eeq
Since $l$ is odd, the sum is equal to its negative which
shows that it vanishes.  The choice of $l$ is arbitrary - changing 
$l$ shifts the origin of the mother by an even number of steps.  
Since the mother is orthogonal to all integer translates of the scaling 
function, it does not matter where the origin is placed.

This shows that the coefficients $h_l$, satisfying  
\beq
\sum_l h_l =  \sqrt{2}.
\label{eq:iba}
\eeq
\beq
\sum_l h_{l-2m} h_{l} = \delta_{m0} 
\label{eq:ibb}
\eeq
with $g_k$ defined by 
\beq
g_k := (-1)^k h_{l-k} \qquad l \qquad \mbox{odd}
\label{eq:ibc}
\eeq
give a multi-resolution analysis, scaling function, and a mother function. 

The Daubechies order-$K$ wavelets are defined by 
the  
conditions
\beq
\int x^n \psi (x) dx = 0, \qquad n=0,1, \cdots ,K-1.
\label{eq:daubcon}
\eeq
These equations ensure that polynomials of degree $<K-1$ 
can be locally represented by finite linear combinations of scaling 
functions on a fixed scale.  This is a useful property for 
numerical approximations.  The order $K$-Daubechies scaling function has
$2K$ scaling coefficients, with $K=1$ corresponding to the Haar 
wavelets, and each additional value of $K$ adds one more orthogonality 
condition.  

The scaling equation (\ref{eq:iav}) and the moment conditions
(\ref{eq:daubcon}) for the mother wavelet function 
gives the additional
equations necessary to find the Daubechies scaling coefficients, $h_l$:
\[
0 = (x^n, \psi) = (Dx^n, D\psi)
\]
\[
=\int dx x^n 2^{-n-1/2}\sum_m g_m \phi (x-m) .
\]
This gives
\[
\sum_m \int dx (x+m)^n  g_m \phi (x) =0 .
\]
For $n=0$ this gives (using the $n=0$ equation) 
\[
\sum g_m=0,\to     \sum_m (-1)^m h_{l-m} =0,
\]
for $n=1$ this gives
\[
\sum m g_m=0, \to \sum_m m(-1)^m h_{l-m} =0,
\]
for $n=2 \cdots k$ this gives
\[
\sum m^2  g_m=0, \to \sum_m m^2(-1)^m h_{l-m} =0,
\]
\[
\vdots 
\]
\[
\sum m^k  g_m=0, \to \sum_m m^k (-1)^m h_{l-m} =0.
\]
When coupled with 
\[
\sum h_l = \sqrt{2} 
\] 
and the orthonormality constraints,
\[
\sum_l h_l h_{l-2n} = \delta_{n0} 
\]
we get a system of equations that 
can be solved for the Daubechies-$K$ scaling coefficients.  
The cases $K=1,2,3$ have analytic solutions.  These solutions are given in
Table 1.
\begin{center}
\begin{table} 
{\bf Table 1: Scaling Coefficients } \\[1.0ex]
\begin{tabular}{|l|l|l|l|}
\hline
$h_l$ & K=1 & K=2 & K=3   \\
\hline					      		      
$h_0$ &$1 / \sqrt{2}$  & $(1+\sqrt{3})/4\sqrt{2}$ &$(1+\sqrt{10}+\sqrt{5+2\sqrt{10}})/16\sqrt{2}$ \\
$h_1$ &$1 / \sqrt{2}$  & $(3+\sqrt{3})/4\sqrt{2}$ & $(5+\sqrt{10}+3\sqrt{5+2\sqrt{10}})/16\sqrt{2}$ \\
$h_2$ &$0$  & $(3-\sqrt{3})/4\sqrt{2}$ & $(10-2\sqrt{10}+2\sqrt{5+2\sqrt{10}})/16\sqrt{2}$ \\
$h_3$ &$0$  & $(1-\sqrt{3})/4\sqrt{2}$ & $ (10-2\sqrt{10}-2\sqrt{5+2\sqrt{10}})/16\sqrt{2} $ \\
$h_4$ &$0$  & $0$ & $(5+\sqrt{10}-3\sqrt{5+2\sqrt{10}})/16\sqrt{2}$ \\
$h_5$ &$0$  & $0$ & $(1+\sqrt{10}-\sqrt{5+2\sqrt{10}})/16\sqrt{2}$ \\
\hline
\end{tabular}
\end{table}
\end{center}
Scaling coefficients
for other values of $K$ are tabulated in the literature [1].
With the exception of the Haar case ($K=1$), there are two solutions
which are related by reversing the order of the coefficients.
 
Given the scaling coefficients, $h_l$, it is possible to use them to compute
the the scaling function.  While the Fourier transform method can
be used to compute the Haar functions exactly, it is more difficult to
use in the general case.

An alternative is to compute the scaling function exactly on a dense set of
dyadic points.  This construction starts from the scaling equation
in the form:
\beq
\phi (x) = \sum_l \sqrt{2} h_l \phi (2x-l).
\label{eq:ibe}
\eeq
Let $x=n$ to get relations between the values of the scaling function 
at integer points
\beq
\phi (n) = \sum_l \sqrt{2} h_l \phi (2n-l) .
\label{eq:ibf}
\eeq
Set $m=2n-l$ to get 
\beq
\phi (n) = \sum_m \sqrt{2} h_{2n-m} \phi (m)
\label{eq:ibg}
\eeq
This gives the equations
\beq
\phi (n) = \sum_m H_{nm} \phi (m)
\label{eq:ibh}
\eeq
for the non-zero $\phi (n)$ corresponding to $n=1,\cdots, 2K-2$ 
where 
\beq
H_{nm} = \sqrt{2}h_{2n-m} .
\label{eq:ibi}
\eeq 
Eigenvectors of the matrix $H_{mn}$ with eigenvalue 1 are
solutions of the scaling function at integer points - up to
normalization.  

Rather than solve the eigenvalue problem, one of the equations 
can be replaced by the condition 
\beq
\sum_n \phi (n) =1
\eeq
which follows from the assumption that $\int \psi(x) dx =0$. 
(The proof of this statement uses the fact that the translates of the 
scaling function on a fixed scale and the wavelets on all smaller scales is
a basis for square integrable functions.  Since 1 is locally orthogonal 
to all of the wavelets by assumption, 1 can be expressed as a 
linear combination of translates of the scaling function.  The normalization 
condition gives the coefficients of the expansion above.)
The support condition implies that only a finite number of the $\phi (n)$ 
are non-zero.  This condition is independent of the orthonormality
condition.

For the case of the $K=2$ Daubechies wavelets these equations are
\[
\phi (0) = \sqrt{2} h_0 \phi (0)
\]
\[
\phi (1) = \sqrt{2} (h_0 \phi (2) + h_1 \phi (1) + h_2 \phi (0))
\]
\[
\phi (2) = \sqrt{2} (h_1 \phi (3) + h_2 \phi (2) + h_3 \phi (1))
\]
\[
\phi (3) = \sqrt{2} h_3 \phi (3)
\]
\[ 
1 = \phi (0) +  \phi (1) + \phi (2) + \phi (3).
\]
The first and fourth equation give $\phi (0) = \phi (3) =0$ (or 
$h_0=h_1 = 1/\sqrt{2}$ which is the Haar solution).  This also 
follows from the continuity of the wavelets, since $0$ and $3$ are
the boundaries of the support.
The second and third equations are eigenvalue equations
\beq
\left ( 
\begin{array}{c}
\phi (1) \\
\phi (2) 
\end{array}
\right ) = 
\left ( 
\begin{array}{cc}
\sqrt{2} h_1 & \sqrt{2} h_0 \\
\sqrt{2} h_2 & \sqrt{2} h_3 
\end{array}
\right ) 
\left ( 
\begin{array}{c}
\phi (1) \\
\phi (2) 
\end{array}
\right ) .
\eeq
Instead of solving the eigenvalue problem for an eigenvector with 
eigenvalue 1,  use
\beq
\phi (1) + \phi (2) =1
\eeq
with 
\[
\phi (1) = \sqrt{2} (h_0 \phi (2) + h_1 \phi (1))
\]
to get 
\[
\phi (1) = \sqrt{2} (h_0 (1 -\phi (1)) + h_1 \phi (1))
\]
which can be solved for 
\beq
\phi (1) = {\sqrt{2} h_0 \over 1 + \sqrt{2}(h_0-h_1)} 
\eeq
and 
\beq
\phi (2) = {1 - \sqrt{2} h_1 \over 1 + \sqrt{2}(h_0-h_1)} .
\eeq
This gives exact values of the scaling function at 
integer points in terms of the scaling coefficients.  
This solution satisfies 
$\sum_n \phi (n) =1$. In this case there are only two non-zero terms.

In order to construct the scaling function at an arbitrary point
$x$ the first step is to make a dyadic approximation to $x$.  Let
$m$ be an integer that defines a dyadic resolution.  This means that 
we want the dyadic approximation to satisfy the inequality 
$\vert x - x_{approx} \vert < 2^{-m}$.  For any $m$ it 
is possible to find an integer $n$ such that 
\beq
{n \over 2^m} \leq x < {n+1 \over 2^m} .
\eeq
Writing this as
\beq
n \leq 2^m x < n+1 
\eeq
immediately gives
\beq
n := [2^m x] = floor(2^mx)
\eeq
where $[]$ means greatest integer $\leq 2^mx$.

Since the scaling function is continuous, for any $\epsilon >0$ we can find a 
large enough $m$ so 
\[
\vert \phi (x) - \phi ({n \over 2^m} ) \vert < \epsilon .
\]
In what follows we evaluate $\phi (n / 2^m)$ exactly. 
Let $x= n / 2^m$.   We also assume that $0 < n < 2K-1 \times 2^m$, 
otherwise $\phi (x) =0$ by the support condition (in this example 
we consider the case $K=2$) .  In order to evaluate 
$\phi (x)$ note that the scaling equation gives:

\[
\phi (x) = \phi \left ({n \over 2^m} \right ) =
\sqrt{2} D \phi \left ({n\over 2^{m-1}} \right ) =
\]
\[
\sum_{l} \sqrt{2} h_{l_1} T^{l_1} \phi \left ({n\over 2^{m-1}}\right )= 
\sum_{l} \sqrt{2} h_{l_1} \phi \left ({n\over 2^{m-1}}-l_1 \right )
\]
\beq
= 
\sum_{l} \sqrt{2} h_{l_1} \phi \left ({n- 2^{m-1}l \over 2^{m-1}}\right ) 
\eeq
Repeating this process a second time gives
\beq
\phi(x) =
\sum_{l_1,l_2} 2  h_{l_1} h_{l_2}  \phi \left ({n- 2^{m-1}l_1 -
2^{m-2}l_2  \over 2^{m-2}}\right ). 
\eeq
Using the scaling equation $m$ times gives
\beq
\phi(x) =
\sum_{l_1,l_2\cdots l_m} 2^{m/2}  h_{l_1} h_{l_2} 
\cdots h_m  \phi (n- 2^{m-1}l_1 -
2^{m-2}l_2 - \cdots 2l_{m-1}-l_m ) .
\eeq
In this case the last expression is evaluated at integer values 
which gives (for the Daubechies $K=2$ case):
\[
\phi(x) =
\]
\beq
\sum_{l_1,l_2\cdots l_m}  c_{l_1} c_{l_2} 
\cdots c_m \times 
\eeq
\begin{eqnarray}
\left [ \delta_{n- 2^{m-1}l_1 -
2^{m-2}l_2 - \cdots 2l_{m-1}-l_m, 1} 
{\sqrt{2} h_0 \over 1 + \sqrt{2}(h_0-h_1)}
\right . +
\\ 
\left .\delta_{n- 2^{m-1}l_1 -
2^{m-2}l_2 - \cdots 2l_{m-1}-l_m, 2} 
{1 - \sqrt{2} h_1 \over 1 + \sqrt{2}(h_0-h_1)} 
\right ]
\end{eqnarray}
where $c_k := \sqrt{2} h_k$.

This method generalizes to any value of $K$ and any choice of 
scaling coefficients, $h_l$.

The scaling function and mother wavelet for the Daubechies wavelet are
pictured in Figure 1.

\begin{center}
\includegraphics{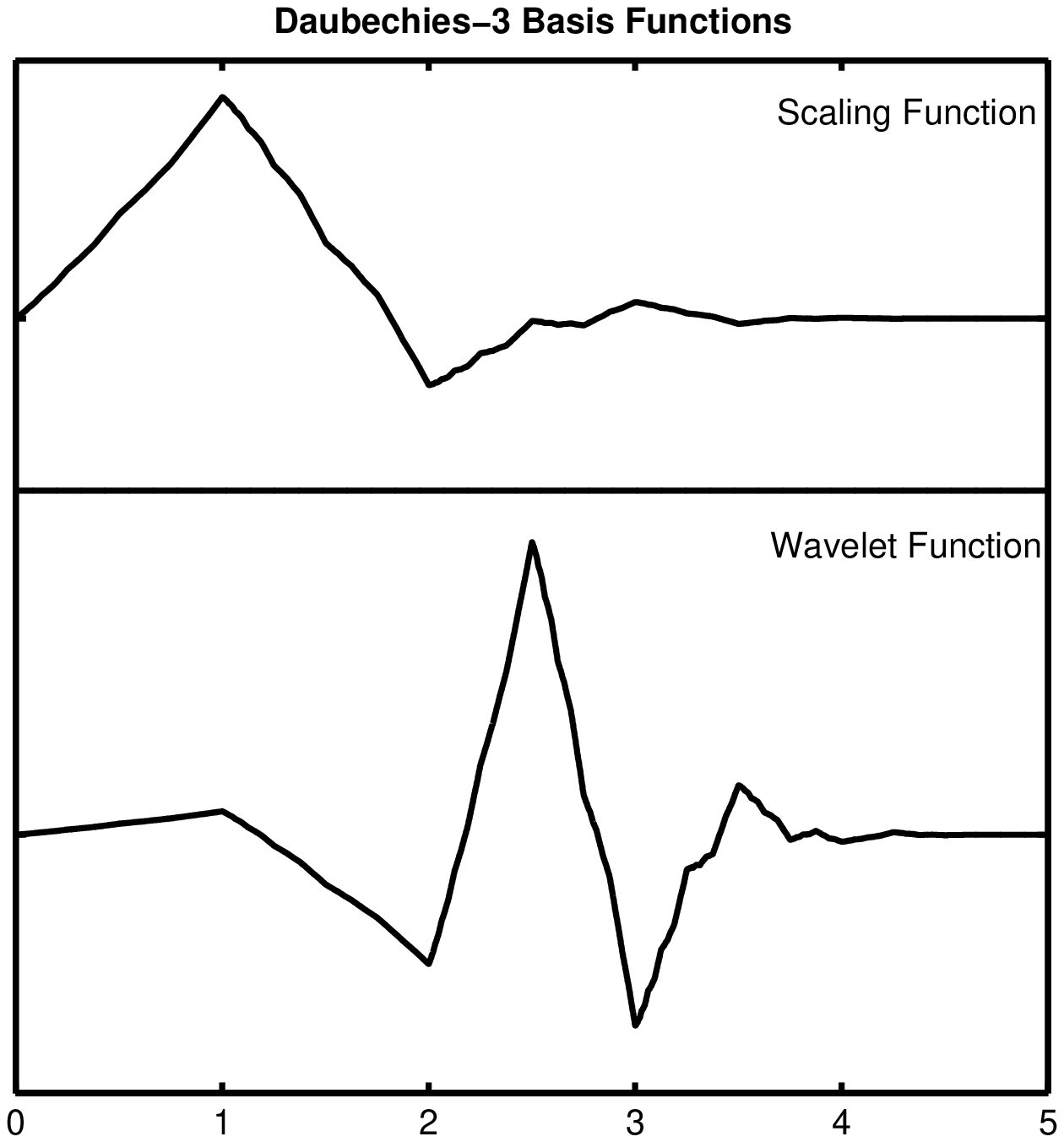}
{\bf Figure 1.}
\end{center}

\section{Daubechies Wavelets}

The Daubechies wavelets have two special properties.  The first is
that there are a finite number of non-zero scaling coefficients, $h_l$,
which means that the scaling functions and wavelets have compact
support.  The order-$K$ Daubechies scaling equation has $2K$ non-zero
scaling coefficients, and the support of the scaling function and
mother wavelet function is on the interval $[0,2K-1]$.  The second property of
the order-$K$ Daubechies wavelets is that the first $K-1$ moments of the
wavelets are zero.

The second property of the Daubechies wavelets is what makes them
useful as basis functions.  The expansion of  a function $f(x)$ in a
wavelet basis has the form
\[
f(x) = \sum_{mn} f_{mn} \psi_{mn}(x) 
\qquad f_{mn} := \int f(x) \psi_{mn} (x) dx .
\] 
If $f(x)$ can be well-approximated by a low-degree polynomial on the
support of $\psi_{mn}(x)$, then the vanishing of the low-order moments
of $\psi_{mn}(x)$ means that the expansion coefficient $f_{mn}$ will
be small.  On the other hand, as we will show in this section, the
scaling function basis can be used to make local pointwise
representation of low-degree polynomials.  Since the scaling function
basis on ${\cal V}_m$ is equivalent to the wavelet basis on all
scales, $k>m$, this means that the wavelet basis provides an efficient
representation of functions that can be accurately approximated by
local polynomials on different scales.  For integral equations with
smooth kernels, this means that the matrix representation of the
kernels in a wavelet basis will be represented by a sparse matrix.

The constraint on the moments of the Daubechies wavelets, 
\beq
\int \psi (x) x^l dx =0 \qquad l=0 \cdots K-1,
\label{eq:wavmom}
\eeq
has important consequences. 
Eq. (\ref{eq:wavmom}) implies  
\[
\int \psi_{0m} (x) x^l dx = \int \psi (x-m) x^l dx = \int \psi (y) (y+m)^l
dy  
\]
\beq
=\sum_{k=0}^l {l! \over k! (l-k)!} m^{l-k} \int \psi (y) y^k dy=0 
\qquad l=0 \cdots K-1 ,
\eeq
which means that first $K-1$ moments of the unit translates of the 
mother wavelet function vanish.
Similarly, changing scale gives
\[
\int \psi_{10} (x) x^l =
\int D \psi (x) x^l dx = {1 \over \sqrt{2}} \int \psi (x/2) x^l dx 
\]
\beq
=2^{l+1/2} \int \psi(y) y^l dy =0 
\qquad l=0 \cdots K-1 .
\eeq
It straight forward to proceed inductively to show for all $m$ and $n$ that  
\beq
\int \psi_{nm} (x) x^l =0 \qquad l=0 \cdots K-1 .
\eeq
This means that every Daubechies wavelet basis function   
is orthogonal to all polynomials of degree less than $K$,
where $K$ is the order of the wavelet basis. 

For the orthonormal basis of $L^2({\mathbb R})$ consisting of
\beq
\lbrace T^n \phi (x) , D^m T^n \psi(x) : m \leq 0 \rbrace
\eeq
the only basis functions with non-zero moments with $l<K$ are the 
scaling basis functions 
\beq
\int \phi_m (x) x^l dx \not= 0 \qquad l=0 \cdots K-1 .
\eeq
Although polynomials are not square integrable, we can multiply 
a polynomial by a box function $b(x)$ which is 1 between $x_-$ and $x_+$
and zero elsewhere.  The product of the box function and the 
polynomial is square integrable and is equal 
to the polynomial on the interval $[x_-,x_+]$.  It follows that 
\beq
p(x)b(x) = \sum_{mn} c_{mn} \psi_{mn} (x) =  
\sum_{n} d_{n} \phi_{kn} (x) + \sum_{n}\sum_{m \leq k} c_{mn} \psi_{mn} (x) 
\label{eq:pbas}
\eeq
where $p(x)$ is a polynomial of degree less than $K$ and 
\beq
c_{mn} = \int_{x_-}^{x_+} \psi_{mn}(x) p(x) dx
\eeq
\beq
d_{n} = \int_{x_-}^{x_+} \phi_{n}(x) p(x) dx .
\eeq
The moment condition means that the coefficients $c_{mn}=0$ whenever
the support of the wavelet is completely contained inside of the
interval $[x_-,x_+]$.  Thus in the first expression the non-zero
coefficients arise from end-point contributions and from many small
contributions from wavelets with support that are much larger than the
box.

If $k$ is set to correspond to a sufficiently fine scale, so the
support of all of the wavelets is much smaller than the support of the
box, then the second sum in (\ref{eq:pbas}) has no wavelets with
support larger than the width of the box.  The endpoint contributions
only affect the answer within a distance $\Delta$, equal to the width
of the support of the scaling basis function, from the endpoints of
the box.  Inside this distance the only nonzero coefficient are due to
the translates of the scaling functions.  There are a finite number of
these coefficients, and in this region they provide an exact
representation of the polynomial.  Specifically let
\beq
I(x) = b(x) p(x) - \sum_{n} d_{n} \phi_{kn} (x) + \sum_{n}\sum_{m \leq k} 
c_{mn} \psi_{mn} (x)
\eeq
then we have 
\[
0= \Vert I \Vert^2 =
\int_{x_-}^{x_- + \Delta} I(x)^2 dx + \int_{x_+-\Delta}^{x_+} I(x)^2 dx + 
\]
\beq
\int_{x_- + \Delta}^{x_+-\Delta}  
\vert p(x)-\sum_{n} d_{n} \phi_{kn} (x)\vert^2 dx .
\eeq
Since all three terms are non-negative we conclude that 
\beq
\int_{x_- + \Delta}^{x_+-\Delta}  
\vert p(x)-\sum_{n} d_{n} \phi_{kn} (x)\vert^2 dx = 0 .
\eeq
Since $\Delta$ is fixed by the choice of the support of the 
scaling function and $x_{\pm}$ is 
arbitrary we have 
\beq
\int_{a}^{b}  
\vert p(x)-\sum_{n} d_{n} \phi_{kn} (x)\vert^2 dx = 0
\eeq
for any interval $[a,b]$.  Since $p(x)$ and $\phi (x)$ are continuous 
(we did not prove this for $\phi(x)$) and the sum of translates 
has a finite number of non-zero terms, it follows that 
\beq
p(x)=\sum_{n} d_{kn} \phi_{n} (x)  
\eeq
pointwise on every finite interval.  Since the box support is arbitrary this
holds for any $k$.  This establishes the desired result, that polynomials 
of degree less than $K$ can be represented exactly by the finite linear 
combinations of the scaling functions $\phi_n (x)$.  Since both bases 
in (\ref{eq:pbas}) are equivalent, it follows that local polynomials
can also be represented exactly in the wavelet basis.

Figure 2. shows integer translates of the Daubechies 2 scaling function.  
Note how the sum of the non zero wavelets at any point in identically one, 
in spite of the complex fractal nature of each individual scaling function.

\begin{center}
\includegraphics{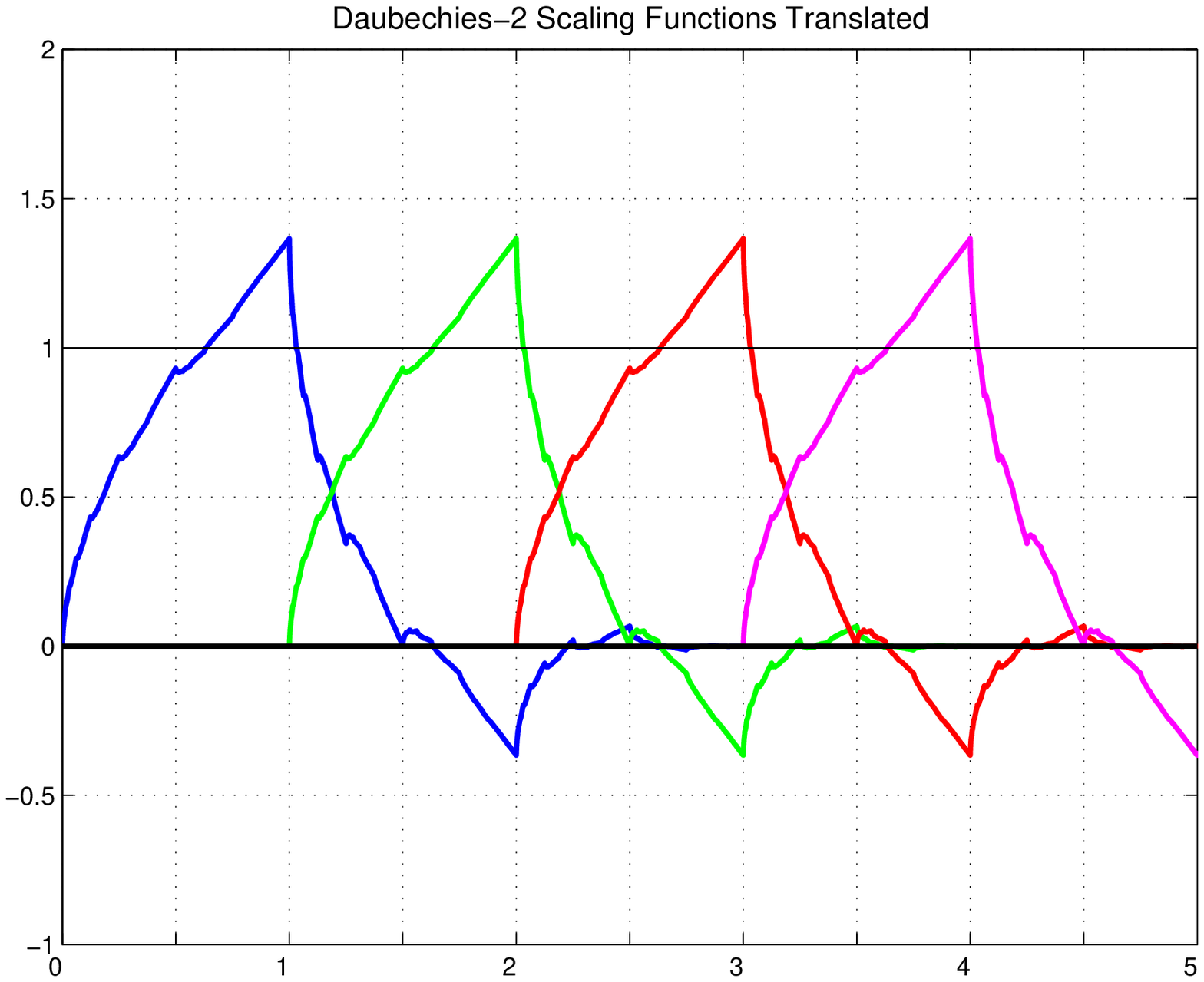}
{\bf Figure 2.}
\end{center}

Figure 2. shows a local representation of a constant function in terms of 
scaling fucntions. 

\begin{center}
\includegraphics{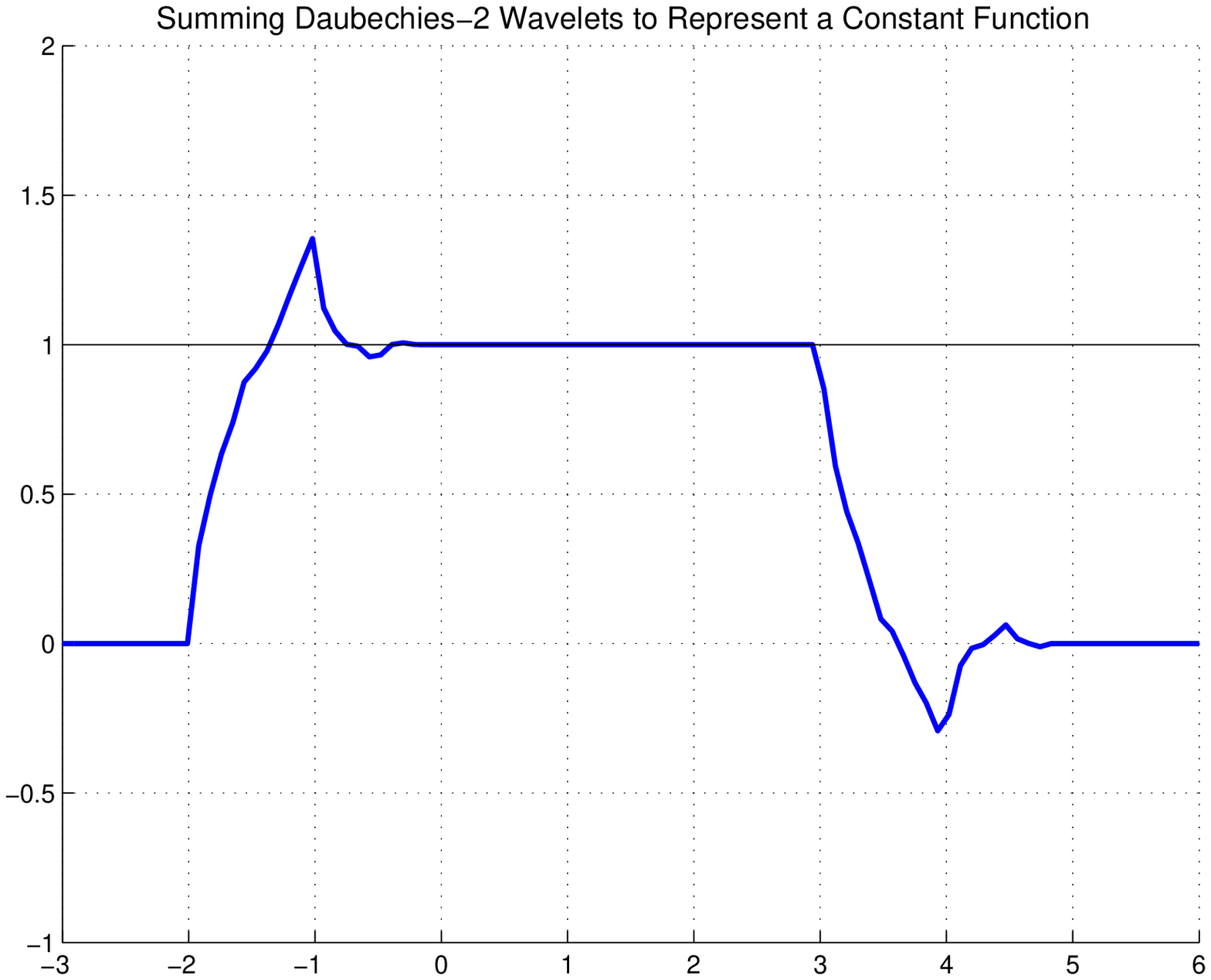}
{\bf Figure 3.}
\end{center}

Figure 3. shows a local representation of a linear function in terms of 
scaling functions, while Figure 4. shows a local representation of a 
linear function.

\begin{center}
\includegraphics{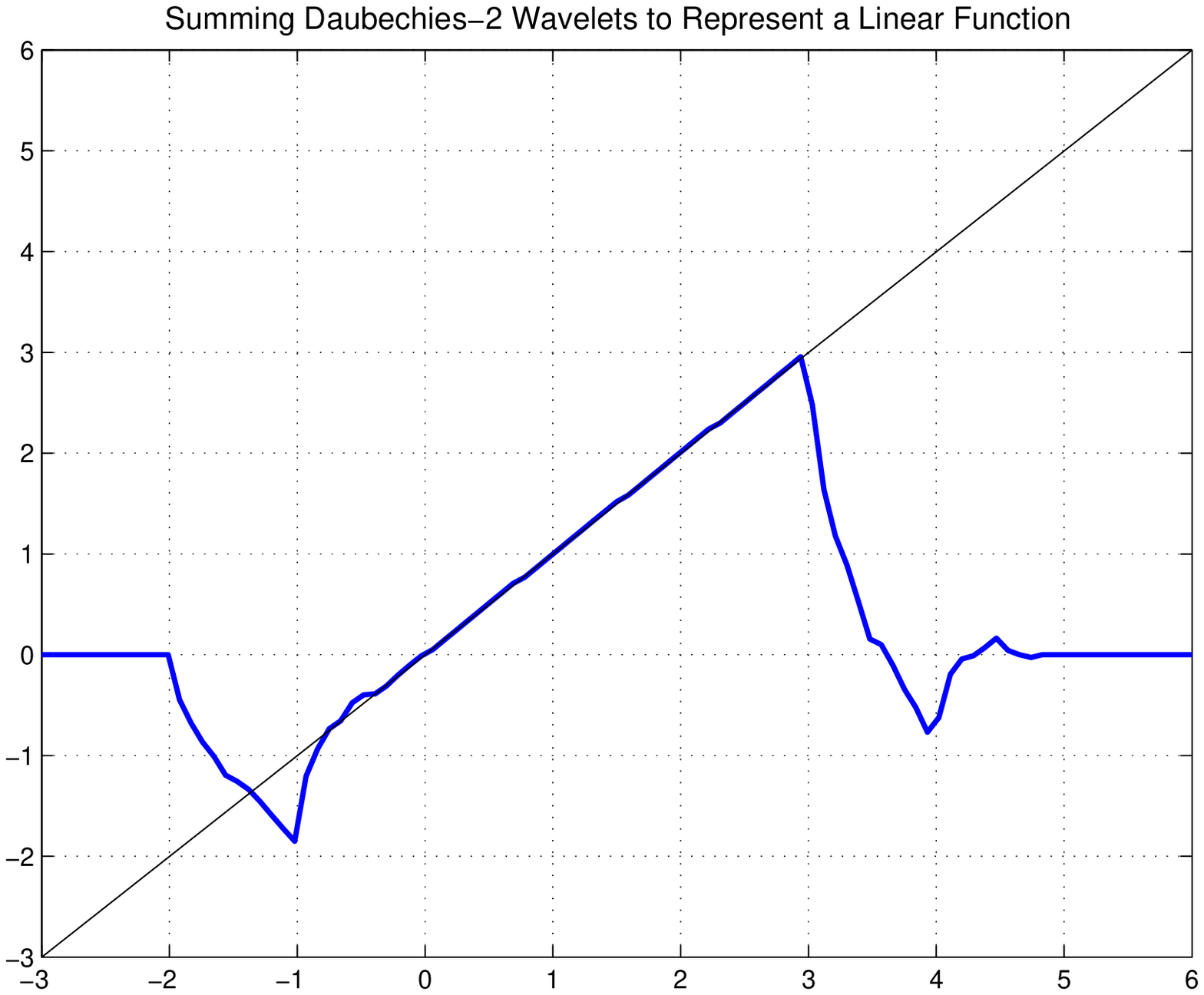}
{\bf Figure 4.}
\end{center}
  
Note that expansion in the wavelet basis gives all coefficients zero.  This is
not a contradiction because none of the polynomials are square integrable.
The key point is that once 
one puts a box around a function,  wavelets with very large support 
(large m) lead to many small contributions.

\section{Moments and Quadrature Rules}

One of the most important properties of the scaling equation is that it
can be used to generate linear relations between integrals involving
different scaling-function or wavelet basis elements.  In this section
we show how the scaling equation can be used to obtain exact
expressions for moments of the scaling function and wavelet basis
elements as functions of the scaling coefficients.  These can be used
to develop quadrature rules that can be applied to linear integral
equations.  In section 6 the scaling equation is used to obtain exact
expressions for the inner products of the these functions and their
derivatives, which are important for applications to differential
equations.  We also show that these same methods can be used to
compute integrals of scaling functions and wavelets over the different
types of integrable singularities encountered in singular integral 
equations.

{\bf Moments} of the scaling function and mother wavelet function 
are defined by
\beq
<x^m>_{\phi} = \int \phi (x) x^m dx 
\qquad
<x^m>_{\psi} = \int \psi (x) x^m dx .
\eeq
Normally these are integrated over the real line. For compactly supported 
wavelets this is equivalent to integrating over the support of the wavelet.  

A polynomial {\bf quadrature rule} is a collection of $N$ points $\{x_i\}$ 
and weights $\{ w_i \}$ with the property 
\beq
<x^m>_{\phi} = \int \phi (x) x^m dx = \sum_{i=1}^N x_i^m w_i
\eeq
which hold for $0\leq m \leq 2N-1$.  By linearity this means that 
\beq
\int \phi (x) P(x) dx = \sum_{i=1}^N P(x_i) w_i
\eeq
is exact for all polynomials of degree up to $2N-1$.

In order to construct a quadrature rule we need to first compute
the moments,  and from these we can compute the points and weights.
The moments can be constructed recursively from the normalization
condition 
\beq
<x^0>_{\phi} = (x^0, \phi) = \int dx \phi (x) = 1
\label{eq:reca}
\eeq
using the scaling equation
\[
<x^m>_{\phi} = (x^m ,\phi )  = (Dx^m ,D\phi ) 
\]
\[
={1 \over \sqrt{2}}{1 \over 2^m} \sum_l h_l (x^m , T^l \phi)   
\]
\[
={1 \over \sqrt{2}}{1 \over 2^m} \sum_l h_l ((x+l)^m , \phi) 
\]
\[
={1 \over \sqrt{2}}{1 \over 2^m} \sum_l h_l \sum_{k=0}^m
{ m! \over k! (m-k)!} l^{m-k} <x^k>_{\phi} .
\]
Using $\sum_{l} h_l = \sqrt{2}$, and moving the $k=m$ term to the left side 
of the above equation gives the recursion relation:
\beq
<x^m>_{\phi} =  {1 \over 2^m-1}{ 1 \over \sqrt{2}}  \sum_{k=0}^{m-1}
{ m! \over k! (m-k)!} \left (\sum_{l=1}^{2K-1} h_l l^{m-k}\right ) 
<x^k>_{\phi}. 
\label{eq:recb}
\eeq
Note that the right hand side of this equation involves moments with $k<m$.
Similarly the moments of the mother wavelet function are expressed in 
terms of the moments of the scaling functions using eq. (\ref{eq:iav}) 
\beq
<x^m>_{\psi} = {1 \over 2^m}  \sum_{k=0}^{m} { m! \over k! (m-k)!} 
\left ( \sum_{l=0}^{2K-1} {g_l \over \sqrt{2}} l^{m-k}\right ) <x^k>_{\phi}. 
\label{eq:recc}
\eeq
Since the scaling equation for the mother wavelet function relates the mother 
wavelet function to the scaling 
function there is no need to take the $k=m$ term to the left of the 
equation; it is known from the first recursion.

Equations (\ref{eq:reca}), (\ref{eq:recb}), and (\ref{eq:recc}) give a
recursive method for generating all non-negative moments of the
scaling and mother wavelet function from the normalization integral of
the scaling function.

The moments for $\phi_{kl} = D^k T^l \phi $ and 
$\psi_{kl} = D^k T^l \psi$ can be computed from 
the moments (\ref{eq:recb}) and (\ref{eq:recc}) using 
the unitarity of the $D$ and $T$ operators 
\[
<x^m>_{\phi_{kl} } = (x^m ,D^k T^l \phi ) = 
(T^{-l} D^{-k} x^m , \phi ) 
\]
\[
= 2^{k(m+1/2)}\sum_{n=0}^m {m! \over n! (m-n)!} l^{m-n} <x^n>_{\phi}   
\] 
and
\[
<x^m>_{\psi_{kl} } = (x^m ,D^k T^l \psi ) = 
(T^{-l} D^{-k} x^m , \phi ) 
\]
\[
=2^{k(m+1/2)}\sum_{n=0}^m {m! \over n! (m-n)!} l^{m-n} <x^n>_{\psi} .  
\] 
Thus, all moments of translates and scale transforms of
both the mother wavelet and scaling functions can be 
computed exactly in terms 
of the scaling coefficients.

\bigskip
\noindent {\bf Partial Moments:}
\bigskip
Partial moments of the form
\[
<x^m>_{\phi [0:n]} = \int_0^n \phi(x) x^m dx 
\]
and
\[
<x^m>_{\phi [n:2K-1]} = \int_n^{2K-1} \phi(x) x^m dx =
<x^m>_{\phi}- <x^m>_{\phi [0:n]} 
\]
for $n \in \{1, \cdots , 2K-2 \}$ are also needed of numerical 
applications.  

These are can be calculated recursively in terms of the full moments
using the scaling equation.  
First consider the order $m=0$ 
partial moments.  Use the scaling equation
\[
D \phi (x) = {1 \over \sqrt{2} } \phi ({x \over 2}) = 
\sum_l h_l \phi (x-l) ,
\]
which gives
\[
\phi (x) = \sum_l \sqrt{2} h_l \phi (2x-l),  
\]
in the definition of the $m=0$ partial moment to obtain:
\[
<x^0>_{\phi [0:n]} = \int_0^n \phi(x) dx = 
\sum_l \sqrt{2} h_l \int_0^n  \phi(2x-l) dx . 
\]
Substituting $y=2x-l$ gives 
\[
<x^0>_{\phi [0:n]} =  
\sum_l  {h_l \over \sqrt{2}} \int_{-l}^{2n-l}   \phi(y) dy =
\sum_l  {h_l \over \sqrt{2}} <x^0>_{\phi [-l,2n-l]}.
\]
The support condition of the scaling function $\phi (x) $ 
implies that the lower limit 
of all of the integrals can be taken as zero which gives the relations:
\[
<x^0>_{\phi [0:n]} =  
\sum_{k=2n-2K+1}^{2n}  c_{2n-k} <x^0>_{\phi [0:k]}
\]
where 
\[
c_l := {h_l \over \sqrt{2}}
\]
are non-zero for $l=0,\cdots , 2K-1$.
These equations are a linear system for the non-trivial partial $0$-moments,
$m_n = <x^0>_{\phi [0:n]}$, 
in terms of the full $0$-moments $<x^0>_{\phi}=1$.  These equations have 
the form 
\beq
M_{mn} m_n = v_m
\label{eq:jxx}
\eeq
where $n,m: 1 \to 2K-2$ and 
\[
M_{mn} = \delta_{mn} - C_{mn}
\] 
with 
\[
C_{mn}= c_{2m-n} \qquad m< K-1, n=1, \cdots, 2m
\]
\[
C_{mn}= 0 \qquad m< K-1, n=2m+1, \cdots, 2K-2
\]
\[
C_{mn}= c_{2m-n} \qquad m=K-1,K
\]
\[
C_{mn}= c_{2m-n} \qquad m>K,n=2m-2K+1,\cdots,2K-2
\]
\[
C_{mn}= 0 \qquad m>K, n=1,\cdots, 2m-2K
\]
and 
\[
v_n = 0 \qquad n< K
\]
\[
v_n= \sum_{k=0}^{2(n-K)+1}c_k \qquad K \leq n \leq 2K-2 .
\]
The matrix $M:= I-C$ has the form
\[
I-C := \left (
\begin{array}{cccccc}
1-c_1 & - c_0 & 0 & 0 & \cdots & 0\\
-c_3 & 1-c_2 & -c_1 & -c_0 & \cdots 0 \\
\vdots & \vdots &  \vdots &  \vdots &  & 0 \\
-c_{2K-3} & -c_{2K-4} &&\cdots && -c_0 \\ 
-c_{2K-1} & -c_{2K-2} &&\cdots && -c_2 \\ 
\vdots & \vdots &  \vdots &  \vdots &  & \vdots  \\
0 & 0 & \cdots  &0  & -c_{2K-1} & 1- c_{2K-2}  
\end{array} 
\right )
\]
the vector $m_n$, of partial moments,  has the form
\[
m:=  
\left (
\begin{array}{c}
<x^0>_{\phi [0:1]}    \\
<x^0>_{\phi [0:2]}  \\
\vdots \\
<x^0>_{\phi [0:K-1]}   \\ 
<x^0>_{\phi [0:K]}    \\ 
\vdots  \\
<x^0>_{\phi [0:2K-2]}    
\end{array} 
\right ) 
\]
and the driving term has the form
\[
v':= 
\left (
\begin{array}{c}
0 \\
0  \\
\vdots  \\
0 \\ 
c_0+ c_1 \\ 
\vdots  \\
c_0 + \cdots c_{2K-3}   
\end{array} 
\right ). 
\]
The solution of this system of linear equations gives the partial moments
of order zero:
\[
m_n = <x^0>_{\phi [0:n]} = (I-C)^{-1} v .
\]
Complementary partial moments are given by:  
\[
<x^0>_{\phi [n:2K+1]} = 1- <x^0>_{\phi [0,n]}.
\]

Higher order partial moments can be constructed similarly
\[
<x^k>_{\phi [0:n]} := \int_0^n \phi (x) x^k dx 
\]
\[
=\sum_{l=0}^{2K-1} \sqrt{2} h_l \int_0^n \phi (2x-l) x^k dx 
\]
\[
=\sum_{l=0}^{2K-1} { h_l \over 2^k \sqrt{2}}  
\int_{-l}^{2n-l}  \phi (x) (x+l)^k dx 
\]
\[
=\sum_{l=0}^{2K-1} {h_l \over 2^k \sqrt{2} }
\sum_{m=0}^k {k! \over m! (k-m)! } l^{k-m}   
\int_0^{2n-l}  \phi (x) x^m dx .
\]
Rearranging indices, putting terms with the partial moments of the 
highest power on the left gives the following equation for the 
order $k$ partial moments:
\beq
\sum_{r=1}^{2K-2} (\delta_{mr} - {1 \over 2^k} C_{mr} )<x^k>_{\phi [0:r]}
= w_m 
\label{eq:recd}
\eeq
where the inhomogeneous term $w_n$ in (\ref{eq:recd}) is 
\[
w_n = \delta (n \geq K)
\sum_{2n-2K+1}^{r=2n} {c_{2n-r} \over 2^k}   
<x^k>_{\phi} 
\]
\beq
+ \sum_{l=1}^{2K-1} {c_l \over 2^k}
\sum_{m=0}^{k-1} {k! \over m! (k-m)! } l^{k-m}   
<x^m>_{\phi [0:2n-l]} ,
\label{eq:pmoma}
\eeq
which can be expressed in terms of the full moments of order $k$ and 
partial moments of order less than $k$.  
The desired order-$k$ partial moments are obtained by inverting the matrix
\[
<x^k>_{\phi [0:m]}:= \sum_r (\delta_{mr} - {1 \over 2^k} C_{mr} )^{-1} w_r .
\]
Note that $C$ matrix is identical to the $C$ matrix that appears in 
equation for the $0$-order partial moments.
 
Having solved for the partial moments of the scaling function,
it is possible to find the partial moments for the mother 
wavelet function
$<x^m>_{\psi_k [0:n] }$
using
\[
<x^m>_{\psi_k [0:n] }:= {1 \over 2^{m+1/2}}
\sum_l g_l \sum_{k=0}^m {m! l^{m-k} \over k! (m-k)!} 
<x^m>_{\phi_k [0:2n-l]}
\]
where the $<x^m>_{\phi_k [0:2n-l]}$ vanish for $2n-l\leq 0$, are
partial moments for $0 < 2n-l < 2K-1$, and are full moments for $2n-l
\geq 2K-1$.  This equation expresses the partial moments of the mother
wavelet function directly terms of moments and partial moments of the
scaling function.

Given the moments and partial moments of $\phi (x)$ and $\psi (x)$ we
can solve for the partial moments of $\phi_{mn}$ and $\psi_{mn}$ in
terms of the partial moments of $\phi (x)$ and $\psi(x)$
by rescaling and translation.  

\bigskip

\noindent{\bf Quadrature Rules:} Given exact the expression for the
moments it is possible to formulate quadrature rules for integrating
the product smooth functions times wavelet or scaling basis functions.
The simplest quadrature rule is the {\bf one-point quadrature rule}. To 
understand this rule consider integrals of the form
\beq
\int f(x) \phi (x) dx .
\eeq
The quadrature point is defined as the first moment of the scaling function
\beq
x_1 := \int x \phi (x) dx= \mu_1 .
\eeq
With this definition we have
\beq
\int (a+bx) \phi (x) dx = a + b \mu_1 = a+b x_1 .
\eeq
For orthogonal wavelets note that 
\[ 
k_m := \int x \phi (x) \phi (x-m) dx =
\int (y+m)  \phi (y+m ) \phi (y ) dy
\]
\[
= \int x  \phi (x+m ) \phi (x ) dx= k_{-m}. 
\]
It follows that 
\beq
\sum_m  m k_m = \sum_m  (-m) k_m =0 .
\eeq
For the Daubechies order $K>1$ scaling functions
\beq
\sum_m  m \phi (x-m) = x- \mu_1  
\eeq
which gives
\beq
0 = \sum_m m  \int x \phi (x) \phi (x-m) dx = 
\sum_m  \int  \phi (x) x ( x-\mu_1 ) dx = \mu_2 - \mu_1^2 . 
\eeq
This means that $\mu_2 = \mu_1^2$ or 
\beq  
\int \phi (x) (a+bx +cx^2 )dx = a + b \mu_1 +c \mu_2 = a + b x_1 + c x_1^2 .
\eeq
This gives a one-point quadrature rule with point $x_1 = \mu_1$ and 
weight $w_1= 1$, that integrates the product of the scaling function 
and polynomials of degree 2 exactly.  This is very useful and simple
when used with scaling basis functions that have small support.

More generally, given a collection of $2N$ moments of the scaling
function or mother wavelet we can construct 
quadrature points and weights using the following method
\cite{payne}.  If $\{ x_i \}$ are the (unknown) quadrature 
points define the polynomial 
\[
P(x) = \prod_{i=1}^N (x -x_i) = \sum_{n=0}^N p_n x^n
\]
where $p_N =1$ and the other $p_n$'s are unknown.  Define 
the polynomials of degree $n+m$
\[
Q_m (x) = x^m P(x) 
\]
for $m=1, \cdots N-1$.  By construction,  for each $m$ and $x_i$ , 
$Q_m (x_i) =0$ because $P(x_i)=0$.  

If we require that the points $\{ x_i \}$ and weights 
$\{ w_i \}$ exactly reproduce $2N$ moments 
then it follows that 
\beq
\int \phi (x) Q_m (x) dx = \sum_{i=1}^N Q_m (x_i) w_i =0
\label{eq:quada}
\eeq
because $Q_m (x_i) =0$.  The condition that the weights reproduce the 
moments give the conditions 
\[
\int \phi (x) Q_m (x) dx = 
\sum_{n=0}^N p_n <x^{n+m}>_{\phi},
\label{eq:quadaa}
\]
and this must be equal to zero for each value $m$ from $m=0$ to $m=N-1$. This 
gives $N$ linear equations for the $N$ unknowns $p_0 \cdots p_{N-1}$:
\[
\sum_{n=0}^N p_n <x^{n+m}>_{\phi} =0 \qquad m=1 \cdots N; \, p_N=1
\]
or
\[
\sum_{n=0}^{N-1}  <x^{n+m}>_{\phi} p_n  = - <x^{N+m}>_{\phi} p_N
\qquad m=1 \cdots N; \, p_N=1 .
\]
Solving this linear system for the coefficients $p_n$  
, using $p_N=1$, gives the polynomial $P(x)$.

Given the polynomial $P(x)$ the next step is to find its zeros.  
The $N$ zeros of the polynomial $P(x)$ are 
the quadrature points $x_i$.  Given 
the quadrature points, the weights are
determined from the remaining $N$ moments by solving the linear system
\[
<x^n>_{\phi} = \sum_{i=1}^N  x_i^n w_i \qquad n=0, \cdots , N-1
\]
for the weights, $w_i$.

This shows how to construct the quadrature points and weights from the
moments.  In applications the linear equations for the coefficients $p_n$
are real .  It follows that the zeros of $P(x)$ are either real or
come in complex conjugate pairs.  In general it is desirable that the
points are real and lie in the support of the scaling function.  When
this fails to occur it is best to simply assign real quadrature points
that lie on the support of the scaling function.  In doing this some
accuracy is sacrificed, but it is easy to go to a higher order.
Generally quadrature rules are used to integrate over the support 
of a scaling function of a small scale;  normally a small number of 
quadrature points and weights is sufficient.

For quadrature rules on a half-interval
the partial moments,
$<x^m>_{\phi_l [0:\infty]}$, need to be used near 0 to generate a
quadrature rule.

Quadrature points are normally needed for different scaling
basis functions, $\phi_{mn}(x)$.  
Points and weights for integrating $\phi_{mn}(x)$ can be generated 
by scaling and translation.
To see this consider a set of points and 
weights $\{ x_i,w_i \}$ that satisfy 
\[
(x^m, \phi ) = \int x^m \phi (x) dx = \sum x_i^m w_i .
\]
It follows that 
\[
(x^m, \phi_{nk} ) = (x^m , D^n T^k \phi ) = 
2^{n(m + 1/2)} (x^m , T^k   \phi , ) 
\]
\[
= 
2^{n(m + 1/2)} ( (x+k)^m, \phi ) = 
\sum 2^{nm + n/2} w_l (x_l+k)^m
\]
\[
 =
\sum (2^{n/2} w_l) (2^n(x_l+k))^m .
\]
If we define the transformed points and weights by 
\[
w'_l = 2^{n/2} w_l \qquad x_l' = 2^n(x_l+k),
\]
we get    
\[
(x^m , \phi_{nk} ) = \sum_l w_l' (x_l')^m .
\]
The new points and weights involve simple transformations of 
the original points and weights.

While it is possible to formulate quadrature rules for both the
wavelet basis and scaling function bases, it makes more sense to
develop the quadrature rules for the scaling function on a
sufficiently fine scale.  This is because the scaling basis functions 
have small support, which means that the quadrature rule will be 
accurate for functions that can be accurately represented by low-degree
polynomials on the support of the scaling function. 

\bigskip 

\noindent{\bf Integral Equations:} To use the quadrature rules to solve 
linear integral equations first consider the non-singular 
integral equation
\[
f(x) = g(x) + \int K(x,y) f(y) dy .
\]
Let 
\[
f(x) \approx \sum_{n} f_n \phi_{sn} (x) 
\]
where $\phi_{sn} (x)$ are translates of the scaling function on a sufficiently 
fine scale $s$.  Inserting the approximate solution in the integral equation
gives 
\[
\sum_{n} f_n \phi_{sn} (x) \approx  g(x) + 
\sum_n \int K(x,y) f_n \phi_{sn} (y)  dy .
\]
Using the orthonormality of the $\phi_{sm}(x)$ for different 
$m$ values and a suitable quadrature rule gives the equation for the 
coefficients $f_m$:
\[
f_m = \sum_l g(x_{lm})w_{lm}  + 
\sum_n \sum_{l,k} w_{lm}  K(x_{lm},x_{kn}) w_{kn} f_n
\]
or
\beq
\sum_n \left [ \delta_{mn} - \sum_{l,k} w_{lm}K(x_{lm},x_{kn})w_{kn}
\right ] f_n
= \sum_l g(x_{lm})w_{lm} .
\label{eq:ieq}
\eeq
Note that no integrals need to be evaluated, except using the local quadrature 
rules.  In addition the points and weights only have to be calculated 
for the scaling function on one scale - the rest can be obtained by simple 
transformations.  

While the scaling function basis on the approximation space ${\cal V}_s$
is useful for deriving the matrix equation above,  it is useful to use
the multiresolution analysis to express the approximation space as
\[
{\cal V}_s = {\cal W}_{s+1} \oplus {\cal W}_{s+2} \oplus \cdots
\oplus {\cal W}_{s+r} \oplus {\cal V}_{s+r} . 
\]
The representation on the right has a natural basis consisting of
scaling basis functions $\phi_{s+r,m}(x) $ on the larger scale, $s+r >
s$, and wavelet basis functions $\psi_{mn}(x)$ on intermediate scales
$s<m\leq s+r$.  These two bases are related by a real orthogonal
transformation called the {\bf  wavelet transform}.  Normally the wavelet
transform is an infinite matrix.  In applications it is replaced by finite
orthogonal transformation that uses some conventions about how to
treat the boundary.

To solve the integral equation the last step is to use the wavelet
transform on the indices $m,n$.  This should give a sparse linear
system that can be used to solve for $f_n$.  While the precise form of
the sparse matrix will depend on how the boundary terms are treated,
the solution in the space ${\cal V}_m$ is independent of the choice of
orthogonal transformation used to get the sparse-matrix form of the 
equations.

Given the solution $f_n$ it is possible to construct $f(x)$ for 
any $x$ using the interpolation
\[
f (x)  = g(x)   + 
\sum_n \sum_{k}  K(x ,x_{kn}) w_{kn} f_n .
\] 
This method has the feature that the solution can be obtained without
ever evaluating a wavelet or scaling function.   The wavelet 
nature of this calculation appears in the quadrature points and
weights, which are functions of the scaling coefficients.

Scattering integral equations have two complications.  First the integral 
is over a half line.  Second, the kernel has a $1/(x\pm i\epsilon)$
singularity.

The endpoint near $x=0$ of the half line can be treated using 
special quadratures for the 
functions on the half interval.  If there is a $\phi_n$ with support 
containing an endpoint, the 
$\delta_{mn}$ in (\ref{eq:ieq}) needs to be replaced by
\[
\int_0^\infty \phi_m (x) \phi_n (x) dx = 
N_{mn} = N_{nm},
\]
which is not a Kronecker delta when the support of $\phi_m$ and 
$\phi_n$ contain the origin.
These integrals can be evaluated using the same methods that 
were used to calculate moments on the half interval.  We use the 
scaling equations and the orthonormality when the support of both terms 
are in the half interval. Specifically for $a$ and $b$ integers
\[
N^{a,b}_{i,j} = \int_a^b \phi_i (x) \phi_j (x) dx 
\]
\[
=\int_{a-i}^{b-i} \phi (x) \phi (x+i-j) dx  
\]
\[
=\sum_{l,l'} h_l h_{l'} \int_{a-i}^{b-i} \phi (2x-l) \phi  (2x+2i-2j-l') 2 dx 
\]
\[
=\sum_{l,l'} h_l h_{l'} \int_{2(a-i)-l}^{2(b-i)-l} \phi (x) 
\phi  (x+2i-2j+l-l')  dx 
\]
\[
=\sum_{l,l'} h_l h_{l'} N^{2a-2i-l,2b-2i-l)}_{0,-2i+2j-l+l'} .
\]
When either function has support inside the interval this is a Kronecker 
delta.  These equations are linear equations that relate these known 
elements to the unknown elements where 
the support overlaps an upper or lower endpoint.  These formulas 
simplify if one of the endpoints satisfy $a= \infty$ or $b=-\infty$.  
The final relations are 
\[
N^{a-i,b-i}_{0,j-i} 
=\sum_{l,l'} h_l h_l' N^{2a-2i-l,2b-2i-l}_{0,-2i+2j-l+l'}.
\]
Note that $N_{0,k} = 1$ if $k=0$, $N_{0,k}=0$ if $k>0$ or $k\leq -(2K-1)$.
It is non-trivial for $-(2K-2)\leq k < 0$.  This gives a linear system for 
the overlap coefficients, $N_{i,j}$. 

For scaling functions that 
overlap $x=0$ the equation becomes: 
\[
\sum_n \left [ N_{0,n-m} - \sum_{l,k} \tilde{w}_{lm}K(\tilde{x}_{lm},\tilde{x}_{kn})
\tilde{w}_{kn}\right ]  f_n
= \sum_l g(\tilde{x}_{lm})\tilde{w}_{lm}
\]
where the $\tilde{x}_{lm},\tilde{w}_{lm}$ indicate that 
for m satisfying $2K-2\leq m <0$ the quadrature points and weights need 
to be replaced by the ones for the half interval.  In this case
overlap matrix elements $N_{mn}$ need to be computed on the scale
dictated by the approximation space ${\cal V}_k$.

Mapping techniques are valuable for transforming the equation to an
equivalent equation on a finite interval and for treating singularities.
For example, the mapping 
\[
x = -x_0 {b \over a}{y-a \over b-y}
\]
transforms the domain of integration to $[a,b]$ and a singularity 
at $x=x_0$ to the origin.
What remains is a mechanism for treating an integrable singularity.
The first step is to use a mapping, like the one above,  
to place the singularity at the origin.  After mapping, 
the relevant integrals for a $1/(x-x_0)$ singularity,
when using the subtraction technique discussed below,
are 
\[
I_m(n) := \int {D^m T^n \phi (x) \over x} dx .
\]
Using unitarity of $D$ gives
\[
I_m(n) := \int D(D^m T^n \phi (x))D{1 \over x} dx =
\]
\[
{2 \over \sqrt{2}} \int {D^{m+1} T^n \phi (x) \over x} dx =
\]
\[
\sqrt{2} \int {D^{m} T^{2n} D \phi (x) \over x} dx =
\]
\[
\sum_{l=0}^{2K-1} \sqrt{2} h_l I_m(2n+l) .
\]
The equations
\[
I_m(n) = \sum_{l=0}^{2K-1} \sqrt{2} h_l I_m(2n+l) 
\]
give linear equations relating the integrals with singularities to
the integrals with no singularities.  The singular terms for the 
order-K Daubechies scaling functions are  
\[
\phi_{m0}(-1), \phi_{m0}(-2), \cdots \phi_{m0}(-2K+2)
\]
The endpoint terms, $\phi_{m0} (0)$ and $\phi_{m0} (-2K+1)$ are not singular because 
$\phi_m (x)$ must be continuous at the endpoints.

We found that these equations are ill-conditioned, but  they can be 
supplemented by 
\beq
0 = \sum_n {\cal P} \int_{-k}^{k} {\phi_{mn} (x) \over x} dx ,
\label{eq:pva}
\eeq
which has the form 
\beq
0 = \sum_n I_{m}^k(n)
\label{eq:sing}
\eeq
where the integrals $I_m^k(n)$ include partial integrals when 
$n$ is such that the support of $\phi_{mn}(x)$ contains the 
points $k$ or $-k$.   
These linear relations relate the singular integrals to the non-singular
ones.  For $\phi_{mn}(x)$ with support far enough away from the origin and 
not containing $k$ or $-k$ the integrals can be expressed in 
terms of the moments:
\[
I_m(n) = \int {D^m T^n \phi (x) \over x} dx = 
\int  T^n \phi (x)  D^{-m}{ 1\over x} dx =
2^{-{m \over 2}} \int \phi (x) T^{-n} { 1\over x} dx 
\]
\[
=2^{-{m \over 2}} \int \phi (x) { 1\over x+n } dx =
2^{-{m \over 2}} {1 \over n} \sum_{l=0}^{\infty} \left ( -{1\over n} \right )^l
\langle x^l \rangle_{\phi} . 
\]
For large values of $n$ this series converges rapidly.  Similar
methods can be used for values of $n$ where $k$ or $-k$ is in the
support of $\phi_{mn}(x)$.  In this case the full moments need to be
replaced by the appropriate partial moments.

For singularities of the form $1/ (x \pm i \epsilon)$  
equation (\ref{eq:pva}) is replaced by 
\[
\int_{-m}^{m} {dx \over x\pm i 0^+} =  \mp i \pi .
\]
Using this with the wavelet expansion 
\[
\sum_n \phi(x) = 1
\]
provides the needed additional equation,
\[
\mp 2 \pi i  = \sum_n I_{m}^k(n) ,
\]
which replaces (\ref{eq:sing}). 
The result of solving these linear equations is accurate approximations to 
the integrals 
\beq
\int {\phi_n (x) \over x \pm i0^+} .
\eeq

To use these to solve the integral equation consider the case $m=0$: 
\[
\int  {K(x,y) \over y} \phi_n(y) dy
\]
\[
=  
\int  {K(x,y)- K(x,0)\over y} \phi_n(y) dy + 
K(x,0)\int {\phi_n(y) \over y}  dy 
\]
\[
=\sum_l {K(x,x_l)- K(x,0)\over x_l} w_l  +
K(x,0)I_0 (n) 
\]
In applications the $I_0(m)$ should be computed for $m$ values
far from the singularity using the series expansion.  The equations 
relating the $I_0(m)$ are used to calculate the remaining $I_0(n)$'s.

Similar methods can be used to treat other types of integrable 
singularities.  For example, for logarithmic singularities define 
\[
I_0(n) := \int \phi_n (x) \ln (x) dx  
\]
The scaling equation gives the linear relations
\[
I_0(n) = (T^n \phi , \ln ) =  (DT^n \phi , D\ln )
\]
\[
={1 \over \sqrt{2}} [(T^{2n} D \phi ,\ln) - \ln (2) (T^{2n} D \phi ,1)] 
\]
\[
=
{1 \over \sqrt{2} }\left ( \sum_l h_l I_0(2n+l) - \ln (2) \right ).  
\]
In this case, because the singularity is integrable and the value of 
the integral is unambiguous, we do not need an additional equation
to specify the treatment of the singularity; however the function is
multiply valued, so the computation of the input integrals far from the
singularity should reflect the choice of Riemann sheet. 
 
The linear equations above relate the integrals $I_0(n)$ that overlap the 
singularity to integrals far away from the singularity, which may or 
may not have support containing the endpoints $\pm a$.  These terms 
serve as input to the linear system and can be computed with the 
moments and partial moments using expansion methods.  For the case
that the support of $\phi_n (x)$ does not contain $\pm a$ the following 
expansion can be used when the support of $\phi_n (x)$ contains 
positive values of $x$:
\[
I_0(n) = \int \phi_n (x) \ln (x) dx = \int \phi (y) \ln (n(1+y/n)) dy
\]
\beq
=\ln (n)- \sum_{m=1}^\infty {(-1)^m \over m}{ <x^m >_{\phi} \over n^m }.
\eeq 
This expansion converges rapidly for large $n$. When $\vert n \vert $ is
large with $n<0$ we use 
\[
\ln (n) = ln \vert n \vert + i (2m+1) \pi
\]
and $m$ is fixed by the treatment of the integral near the origin.
In the case that one
of the endpoints $\pm a$ are contained in the support of $\phi_n$ the
moments, a similar expression can be used to approximate the integrals
$I_0^a(n)$, except the moments, including the $0$-th moment multiplying
$\ln (n)$, need to be replaced by partial moments.

For the case of the Daubechies $K=2$ and $K=3$ wavelets the solution of these
equations in given in Table 2.  The numbers in the table are for  
$\int \phi_n (x) \ln \vert x \vert dx$ for the $\phi_n(x)$ with 
support containing $x=0$.
\begin{center}
\begin{table} 
{\bf Table 2: Singular Integrals } \\[1.0ex]
\begin{tabular}{|l|l|l|}
\hline
$K=2$  &    & \\
\hline		
$n=-2$ & $\int \phi(x-n)\ln \vert x\vert dx$ & $0.456927033732831$ \\
$n=-1$ & $\int \phi(x-n)\ln \vert x\vert dx$ & $-1.64215549088219$ \\
\hline
$K=3$  &    & \\
\hline		
$n=-4$ & $\int \phi(x-n)\ln \vert x\vert dx $ & $ 1.15737952417967$ \\
$n=-3$ & $\int \phi(x-n)\ln \vert x\vert dx $ & $ 0.750468355278047$ \\
$n=-2$ & $\int \phi(x-n)\ln \vert x\vert dx $ & $ 0.315624303943019$ \\
$n=-1$ & $\int \phi(x-n)\ln \vert x\vert dx $ & $ -1.83646456399118$ \\
\hline
\end{tabular}
\end{table}
\end{center}

\section{Derivatives and Differential Equations}

The scaling equation can also be used as a tool to solve differential 
equations.  Consider the following approximation of the function $f(x)$
given by 
\beq
f(x) \sim \sum_{n} f_n \phi_{mn} (x)
\label{eq:expa}
\eeq
where $\phi_{mn}(x) = D^m T^n \phi (x)$ is the scaling function basis on 
the approximation space ${\cal V}_m$.  As $m \to -\infty$ this representation 
becomes exact.

For the purpose solving differential equation we want to calculate 
approximate derivatives of $f(x)$ on the same approximation space
\[
f'(x) := \sum_{n} d'_n \phi_{mn} (x).
\]
\[
f''(x) := \sum_{n} d''_n \phi_{mn} (x).
\]
The orthonormality of the scaling basis functions can be used to 
find the expansion coefficients $d'_n$ and $d_n''$:
\beq
d'_n := \int \phi_{mn} (x) f' (x) dx = - \int \phi'_{mn} (x) f(x) dx. 
\label{eq:expb}
\eeq
\beq
d''_n := \int \phi_{mn} (x) f'' (x) dx =  \int \phi''_{mn} (x) f(x) dx. 
\label{eq:expbb}
\eeq
Using the expansion  (\ref{eq:expa}) in  (\ref{eq:expb}) gives 
\[
d'_n 
= - \sum_{n'}  (\phi'_{mn}, \phi_{mn'}) f_{n'}
\]
and
\[
d''_n 
=  \sum_{n'}  (\phi''_{mn}, \phi_{mn'}) f_{n'} .
\]
The coefficients $(\phi'_{mn} , \phi_{mn'})$ and
$(\phi''_{mn} , \phi_{mn'})$ are needed to compute
the expansion coefficients $d'_n$  and $d_n''$ for the derivatives 
in terms of the expansion coefficients $f_{n'}$ of the function.

Given these linear relations between the coefficients $d_n'$ and
$d_n''$, the solution of a linear differential equation can be reduced
to solving a linear algebraic system for the coefficients $f_n$.  The
size of this system can be reduced by employing the wavelet
transformation.  The method of solution depends on the type of 
problem.  Standard methods can be used to enforce boundary conditions;
the only trick is that all of the basis functions with support that 
overlaps the boundary should be retained.  
 
The goal of this section is to show that the scaling 
equation can be used to compute the needed overlap integrals.

To proceed we first consider the simplest case where the scale $m=0$.
Using unitarity of the translation operator gives the following
relations
\[
(\phi'_{n} , \phi_{n'})= 
(\phi' , \phi_{n'-n})=
(\phi'_{n-n'} ,\phi )
\]
\[
(\phi''_{n}, \phi_{n'})= 
(\phi'' , \phi_{n'-n})=
(\phi''_{n-n'}, \phi ).
\]
In addition these coefficients have the following symmetry relations
\[
(\phi' , \phi_{n})= \int \phi' (x) \phi (x-n) dx   
=- \int \phi (x) \phi'(x-n) dx =
\]
\[
=- \int \phi'(y) \phi_ (y+n)  dy
= -(\phi', \phi_{-n})
\] 
and
\[
(\phi'' , \phi_{n}) = (\phi'', \phi_{-n})
\]
which follow by integration by parts.

The overlap coefficients can be computed using the scaling equation and 
the derivatives of the scaling equation:
\[
\phi (x) = \sqrt{2} \sum_{l=0}^{2K-1}  h_l \phi (2x-l)
\]
\[
\phi' (x) = 2 \sqrt{2} \sum_{l=0}^{2K-1}  h_l \phi' (2x-l)
\]
\[
\phi'' (x) = (2^2) \sqrt{2} \sum_{l=0}^{2K-1}  h_l \phi'' (2x-l).
\]

We first consider the computation of the coefficients
$(\phi' (x), \phi_{n})$.

For $a_n$ defined by 
\[
a_n := (\phi' (x), \phi_{n})
\]
this leads to the following linear relations among the overlap coefficients
\[
a_n = 4 \sum_{l,l'=0}^{2K-1} h_l h_{l'} \int \phi (2x - 2n -l) 
\phi' (2x -l') dx
\]  
\beq
= 2 \sum_{l,l'=0}^{2K-1} h_l h_{l'} \int \phi (y - 2n -l+l') 
\phi' (y )  dy = 
2 \sum_{l,l'=0}^{2K-1} h_l h_{l'} a_{2n +l-l'} .
\label{eq:hom}
\eeq
Since both $\phi (x)$ and $\phi' (x)$ have support for
$0 \leq x \leq (2K-1)$,  the non-zero terms in the sum are constrained by 
\[
-(2K-1) < 2n+l-l' < 2K-1 .
\]
For the second derivative these equations are replaced by 
\beq
a_n := (\phi'' (x), \phi_{n}) =
8 \sum_{l,l'=0}^{2K-1} h_l h_{l'} a_{2n +l-l'} 
\label{eq:homa}
\eeq
These linear equation are homogeneous and must be supplemented by
a normalization condition.  For the Daubechies wavelets of order 
$K>1$ we have the expansion
\[
x = \sum_n b'_n \phi_n(x)  \qquad x^2 = \sum_n b''_n \phi_n (x)
\]
where the expansion coefficients are
\[
b'_n = \int \phi_n (x) x dx = n + \int \phi (x-n) (x-n ) dx
=n + <x>_{\phi}
\]
and
\[
b''_n = \int \phi_n (x) x^2  dx =
\int \phi (x) (x+ n )^2 dx = n^2 + 2n<x>_{\phi} + <x^2>_{\phi}
\]
\[
= n^2 + 2n<x>_{\phi} + <x>^2_{\phi} =
(n+<x>_{\phi})^2 .
\]
Thus
\[
x = <x>_\phi + \sum_n n \phi_n(x) \qquad
x^2 =<x>^2_{\phi} + 2 (x -<x>_\phi ) <x>_\phi + \sum_n n^2 \phi_n (x) 
\]
These equations can be differentiated to get
\[
1 = \sum_n n \phi'_n(x) \qquad
2 = \sum_n n^2 \phi''_n (x)
\]
Multiplying by $\phi (x)$ and integrating gives the desired inhomogeneous 
equation
\[
1 = \sum_n n (\phi , \phi'_n ) =
\sum_n n (\phi_{-n}  , \phi' ) =
\]
\beq
-\sum_n n (\phi_{n}  , \phi' ) = - \sum_n n a_n 
\label{eq:inho}
\eeq
and 
\beq
2= \sum_n n^2 a_n''
\label{eq:inhoa}
\eeq
Equations (\ref{eq:hom}) and (\ref{eq:inho}) or
(\ref{eq:hom}) and (\ref{eq:inhoa}) are linear systems that 
can be used to solve for the coefficients $a_n'$ and $a_n''$.   
  
In general, it is desirable to expand a function using a scaling basis
associated with a sufficiently small scale $m$, In addition, for
efficiency it also useful to use the basis on the approximation space
${\cal V}_m$ consisting or the scaling functions on the scale $m+k$
and wavelet basis functions on scales between $m+1$ and $m+k$.
Finally, one needs to be able to treat higher derivatives.
Generalizations of the methods can be used to find exact expression
for all of these quantities expressed a solutions of linear equations
involving the scaling coefficients.  For the higher derivatives 
it is necessary to use a Daubechies wavelet of sufficiently high
order.  The number of derivative of the wavelet and scaling function 
basis increases with order.  

A necessary condition for the solution of the scaling equation to have 
$k$ derivatives can be obtained by differentiating the scaling equation 
$k$ times, which gives
\[
\phi^{(k)} (x) = \sqrt{2} 2^k \sum_l h_l \phi^{k} (2x-l)
\label{eq:der}
\]
Letting $x=m$ and $n=2m-l$ gives the equation 
\[
\phi^{(k)} (m) = \sqrt{2} 2^k \sum_l h_{2m-n} \phi^{k} (n)
\]
where the non-zero values of $\phi^k (n)$ satisfy $0 \leq n \leq 2K-1$.
For this equation of have a solution the matrix $H_{mn}:= h_{2m-n}$
must have eigenvalue $2^{-(k+1/2)}$.  This is a necessary condition 
for the basis to have $k$ derivatives.   When the $k$-th derivative 
exists, it can be computed up to normalization by iterating the Fourier 
transform of equation (\ref{eq:der}).  The method used to compute 
wavelets at dyadic points can also be used with the above equation to
compute the $k$-th derivatives of scaling functions and wavelets 
at dyadic points.

The scaling equation can be used to exactly compute all of the 
expansion coefficients.  In order 
to exhibit the key relations it is useful to use operators:
\beq
Df(x) = {1 \over \sqrt{2}} f ({x \over 2})
\eeq 
\beq
Tf(x) = f (x -1)
\eeq 
\beq
\Delta f(x) = {df \over dx}(x).
\eeq 

Direct computation shows the following intertwining relations
\beq
\Delta D = { 1 \over 2} D \Delta 
\eeq
\beq
D T = T^2  D  
\eeq
\beq
\Delta T = T \Delta 
\eeq
\beq
\Delta^{\dagger} = - \Delta \qquad T^{\dagger} = T^{-1}
\qquad D^{\dagger} = D^{-1} .
\eeq

We also have the scaling equations:
\beq
D\phi = \sum_l h_l T^l \phi
\eeq
\beq
D\psi = \sum_l g_l T^l \phi .
\eeq

Using the operator relations above give
\beq
D \Delta^r \phi = 2^r \sum_l h_l T^l \Delta^r \phi
\eeq
\beq
D \Delta^r \psi = 2^r \sum_l g_l T^l \Delta^r \phi .
\eeq

The different expansion coefficients can be expressed in terms 
of these operators as
\beq
(\phi_{m'n'} , \Delta^r \phi_{mn}) = ( D^{m'} T^{n'} \phi , 
\Delta^r D^m T^n  \phi)
\eeq
\beq
(\psi_{m'n'}, \Delta^r \phi_{mn} ) = ( D^{m'} T^{n'} \psi, \Delta^r D^m T^n \phi )
\eeq
\beq
(\phi_{k'n'}, \Delta^r \psi_{mn} ) = ( D^{k'} T^{n'} \phi, \Delta^rD^m T^n \psi )
\eeq
\beq
(\psi_{k'n'}, \Delta^r \psi_{mn} )
= ( D^{k'} T^{n'} \psi, \Delta^r D^m T^n \psi ). 
\eeq
The following steps are used to evaluate these coefficients: 

1. Move all of the factors of $D$ to a single side of the equation.  Choose 
the side where the power of $D$ is positive. 

2. Move the $D$'s through all derivatives.

3. Use the scaling equations to eliminate all of the $D$'s.

4. Move all of the $T$'s to the left side of the scalar product.

Using these steps all of the overlap coefficients can be expressed 
in terms of 
\beq
(  \phi_n , \Delta^r  \phi)
\eeq
\beq
( \psi_n  , \Delta^r \phi  )
\eeq
\beq
(\phi_n, \Delta^r \psi,  )
\eeq
\beq
(\psi_n, \Delta^r \psi )
\eeq
The scaling equation can be used to express all of the $\psi$ terms in 
terms of the $\phi$ terms.  The result is at all of the coefficients 
can be expressed in terms of the coefficients 
\beq
( \phi_n , \Delta^r  \phi)
\eeq
We have shown how to compute these for $r=1$ and $2$.  The coefficients for 
larger values of $r$ can be obtained by solving the system:
\[
r! = \sum_n n^{r} \phi^{(r)}_n (x)
\]
\beq
a^{(r)}_n := (\phi'' (x), \phi_{n}) =
2^{2r-1} \sum_{l,l'=0}^{2K-1} h_l h_{l'} a^{(r)}_{2n +l-l'} 
\label{eq:homc}
\eeq

\section{Galerkin for Scattering}
\bigskip

We want to find the solution to the s-wave Schr\"odinger equation
\begin{equation}
-\frac{\hbar^2}{2m}\frac{1}{r}\frac{d^2}{dr^2}[r\,\psi(r)]+V(r)\,\psi(r)
      = E\,\psi(r) \label{scheqn}
\end{equation}
for a particle with mass $m$ and energy $E = \hbar^2 k^2/2m$,
where $\psi(r)$ has the asymptotic form
\begin{equation}
r\psi(r) \,{\lower7pt\hbox{$\displaystyle\longrightarrow$}
    \atop\scriptstyle{r\rightarrow\infty}}\, \sin(kr+\delta) \,,
\end{equation}
and $r\psi(r)$ is zero at the origin. Equation (\ref{scheqn}) can be rewritten
in the form
\begin{equation}
-\frac{1}{r}\frac{d^2}{dr^2}[r\,\psi(r)]+U(r)\,\psi(r)
      = k^2 \psi(r) \,, \label{ueqn}
\end{equation}
where
\begin{equation}
U(r) = \frac{2m}{\hbar^2} V(r) \,.
\end{equation}
To solve Equation (\ref{ueqn}) we choose a complete set of basis functions
$\phi_n(r)$ and write
\begin{equation}
\psi(r) = \sum_{n=1}^N a_n\,\phi_n(r) \,.
\label{psiexp}
\end{equation}
To solve for the expansion coefficients $a_n$ we first
multiply Equation (\ref{ueqn}) by $r^2\phi_m(r)$ and integrate from
$0$ to $R$. This gives
\begin{equation}
-\int_0^R r\,\phi_m(r) \frac{d^2}{dr^2}
     [r\,\psi(r)]\,dr + \int_0^R \phi_m(r)\,U(r)\,\psi(r) r^2dr
     = k^2 \int_0^R \phi_m(r)\,\psi(r) \,r^2dr \label{mproj} \,.
\end{equation}
Using integration by parts, the first term in Equation (\ref{mproj}) can be
written as
\begin{equation}
-\int_0^R r\,\phi_m(r) \frac{d^2}{dr^2}[r\,\psi(r)]\,dr
    = -r\,\phi_m(r) \frac{d}{dr}[r\,\psi(r)]\bigg\vert_0^R
      + \int_0^R \frac{d}{dr}[r\,\phi_m(r)] \frac{d}{dr}[r\,\psi(r)]\,dr \,.
\end{equation}
Now we set
\begin{equation}
\frac{d}{dr}[r\,\psi(r)]\bigg\vert_{r=R} = 1 \label{newbc} \,,
\end{equation}
which corresponds to a change in the normalization of $\psi(r)$, and use
$r\,\psi(r)$ is zero at $r=0$ to write
\begin{equation}
-\int_0^R r\,\phi_m(r) \frac{d^2}{dr^2}[r\,\psi(r)]\,dr
    = -R\,\phi_m(R) + \int_0^R \frac{d}{dr}[r\,\phi_m(r)]
       \frac{d}{dr}[r\,\psi(r)]\,dr  \label{intp} \,.
\end{equation}
Thus, Equation (\ref{mproj}) can be written in the form
\begin{Eqnarray}
\int_0^R \frac{d}{dr}[r\,\phi_m(r)] \frac{d}{dr} [r\,\psi(r)]\,dr
    &+& \int_0^R \phi_m(r)\,U(r)\,\psi(r) r^2dr \nonumber \\
    & & \hspace{-25pt}-\ k^2 \int_0^R \phi_m(r)\,\psi(r) \,r^2dr
     = R\,\phi_m(R) \label{eneqn} \,.
\end{Eqnarray}
Substituting the expansion for $\psi(r)$ in Equation (\ref{psiexp}) into
Equation (\ref{eneqn}) gives
\begin{Eqnarray}
\sum_{n=1}^N a_n \left\{
\int_0^R \right. &\frac{d}{dr}[r\,\phi_m(r)]  \frac{d}{dr} [r\,\phi_n(r)]\,dr 
+ \left. \int_0^R \phi_m(r)\,U(r)\,\phi_n(r) r^2dr \right. \nonumber  \\
& \hspace{-25pt} -\ \left. k^2 \int_0^R \phi_m(r)\,\phi_n(r) \,r^2dr
\right\} = R\,\phi_m(R)  \label{mateqn} \,,
\end{Eqnarray}
which can be written as the matrix equation
\begin{equation}
\sum_{n=1}^N A_{mn}\,a_n = b_m \label{mateqn2} \,.
\end{equation}

Given the solution to Equation (\ref{mateqn2}) we need to determine the
normalization and the phase shift. For the new boundary condition given in
Equation (\ref{newbc})
\begin{equation}
r\psi(r) \,{\lower7pt\hbox{$\displaystyle\longrightarrow$}
    \atop\scriptstyle{r\rightarrow\infty}}\, A\,\sin(kr+\delta) \,, \label{asym}
\end{equation}
Thus, we need an additional equation to use with Equation (\ref{newbc}). We use
the integral
\begin{Eqnarray}
I &=& \int_0^R \sin(kr)\,U(r)\,\psi(r)\,r dr \nonumber \\
  &=& \int_0^R \sin(kr)\left\{\frac{1}{r}\frac{d^2}{dr^2}[r\,\psi(r)]
            + k^2\,\psi(r)\right\}r dr \nonumber \\
  &=& \sin(kr)\frac{d}{dr}[r\,\psi(r)]\bigg\vert_0^R
       -k\cos(kr)[r\,\psi(r)]\bigg\vert_0^R  \label{tmat} \\
  & & \qquad\qquad +\ \int_0^R \left[\left(\frac{1}{r}\frac{d^2}{dr^2}
       + k^2\right)\sin(kr)\right]r\,\psi(r) \,dr \nonumber \\
  &=& kA\left[\sin(kR)\cos(kR+\delta) - \cos(kR)\sin(kR+\delta)\right]
           \nonumber \\
  &=& - kA\sin\delta  \,. \nonumber
\end{Eqnarray}
From Equations (\ref{newbc}) and (\ref{asym}) we get
\begin{equation}
kA\cos(kR+\delta) = kA\cos \delta \cos(kR)-kA\sin \delta \sin(kR) =1 \,,
\end{equation}
which using Equation (\ref{tmat}) can be written as
\begin{equation}
kA\cos\delta = \frac{1-I\sin(kR)}{\cos(kR)} \,. \label{acos}
\end{equation}
Finally, from Equations (\ref{tmat}) and (\ref{acos}) we find
\begin{equation}
\tan \delta = \frac{-I\cos(kR)}{1-I\sin(kR)} \,.
\end{equation}
Given $\delta$ the value of $A$ can be found using Equation (\ref{tmat}).

\medskip
\centerline{\sc Eckart Wave Function}
\medskip
To test the method we use the Eckart potential
\begin{equation}
V(r) = -\frac{\hbar^2}{2m}\frac{2\beta\lambda^2e^{-\lambda r}}
           {(\beta e^{-\lambda r}+1)^2} \,,
\end{equation}
which has the analytic solution
\begin{equation}
r\psi(r) = \frac{\left[\left(4k^2+\lambda^2\right)+(4k^2-\lambda^2)\beta
             e^{-\lambda r}\right]\sin(kr+\delta)-4k\lambda\beta e^{-\lambda r}
             \cos(kr+\delta)}{(4k^2+\lambda^2)\left(\beta e^{-\lambda r}+1
             \right)} \,, \label{scfun}
\end{equation}
where the phase shift is given by
\begin{equation}
\delta = \arctan\left(\frac{\lambda}{2k}\right)+\arctan\left(
           \frac{\lambda(\beta-1)}{2k(\beta+1)}\right) \,. \label{delta}
\end{equation}
If $\beta$ is chosen to have the value
\begin{equation}
\beta = \frac{\lambda+2\kappa}{\lambda-2\kappa} \,
\end{equation}
the potential will have a bound state with the energy
\begin{equation}
E_B = -\frac{\hbar^2\kappa^2}{2m} \,.
\end{equation}
The boundstate wave function is given by
\begin{equation}
r\psi_B(r) = \frac{2\sqrt{2\kappa\beta}\sinh\left(\lambda r/2\right)
               e^{-\kappa r}}{e^{\lambda r/2}
               + \beta e^{-\lambda r/2}} \,.
\end{equation}

\section{Wavelet Filters}
\bigskip
The wavelet transform is the orthogonal mapping between the scaling
function basis on a fine scale and the equivalent basis consisting of
scaling functions on a coarser scale and wavelets at all intermediate
scales.  The wavelet transform can be implemented by treating the
scaling equation and the equation defining the wavelets as linear
combinations of scaling functions for a finer scale, as low and high
pass filters.  This has the advantage when most of the high-frequency
information is unimportant.

The wavelet filter is defined using the scaling relations
\begin{equation}
D\,\phi(x) = \frac{1}{\sqrt{2}}\phi\left(\frac{x}{2}\right)
     = \sum_{l=0}^{2k-1} h_l T^l \phi(x) \label{sc-phi}
\end{equation}
and
\begin{equation}
D\,\psi(x) = \frac{1}{\sqrt{2}}\psi\left(\frac{x}{2}\right)
     = \sum_{l=0}^{2k-1} g_l T^l \phi(x)\,, \label{sc-psi}
\end{equation}
where $g_l = (-1)^l h_{2k-1-l}$.
Using Equations (\ref{sc-phi}) and (\ref{sc-psi}) we can write
\begin{Eqnarray}
\phi_{j,m}(x) &=& D^jT^m \phi(x) \nonumber \\
              &=& \sum_{l=0}^{2k-1} h_l D^j T^m D^{-1} T^l \phi(x) \nonumber \\
              &=& \sum_{l=0}^{2k-1} h_l D^{j-1} T^{2m} T^l\phi(x)
                     \label{sc-phij} \\
              &=& \sum_{l=0}^{2k-1} h_l \phi_{j-1,2m+l}(x) \nonumber
\end{Eqnarray}
\hspace*{-0.28em}and
\begin{Eqnarray}
\psi_{j,m}(x) &=& D^jT^m \psi(x) \nonumber \\
              &=& \sum_{l=0}^{2k-1} g_l D^j T^m D^{-1} T^l \phi(x) \nonumber \\
              &=& \sum_{l=0}^{2k-1} g_l D^{j-1} T^{2m} T^l\phi(x)
                     \label{sc-psij} \\
              &=& \sum_{l=0}^{2k-1} g_l \phi_{j-1,2m+l}(x) \,, \nonumber
\end{Eqnarray}
\hspace*{-0.28em}where we have used
\begin{Eqnarray}
T\,D^{-1}\phi(x) &=& T\,\sqrt{2}\phi(2x) \nonumber \\
                &=& \sqrt{2}\phi(2x-2) \\
                &=& D^{-1} T^2\phi(x) \nonumber
\end{Eqnarray}
\hspace*{-0.28em}to write $T^m D^{-1} = D^{-1} T^{2m}$.

We use the orthonormality of the functions $\phi_{j,m}(x)$ and $\psi_{j,m}(x)$
to obtain the inverse relation. Since ${\cal V}_{j-1} = {\cal V}_j \oplus
{\cal W}_j$, we can write
\begin{equation}
\phi_{j-1,m}(x) = \sum_n a_n \phi_{j,n}(x) + \sum_n b_n \psi_{j,n}(x) \,,
      \label{inv-ab}
\end{equation}
Using the scaling relations, we find
\begin{Eqnarray}
a_n &=& \int \phi_{j,n}(x)\,\phi_{j-1,m}(x)\,dx \nonumber \\
    &=& \sum_l h_l \int \phi_{j-1,2n+l}(x)\,\phi_{j-1,m}(x)\,dx \nonumber \\
    &=& \sum_l h_l \delta_{2n+l,m} \\
    &=& h_{m-2n} \\
\noalign{\hbox{and}}
b_n &=& \int \psi_{j,n}(x)\,\phi_{j-1,m}(x)\,dx \nonumber \\
    &=& \sum_l g_l \int \phi_{j-1,2n+l}(x)\,\phi_{j-1,m}(x)\,dx \nonumber \\
    &=& \sum_l g_l \delta_{2n+l,m} \\
    &=& g_{m-2n} \,.
\end{Eqnarray}
\hspace*{-0.28em}Thus, we find
\begin{equation}
\phi_{j-1,m}(x) = \sum_n h_{m-2n} \phi_{j,n}(x) + \sum_n g_{m-2n} \psi_{j,n}(x)
                    \,. \label{inv-rel}
\end{equation}

An alternate derivation is to use the orthonormality relations
\begin{Eqnarray}
\sum_l h_l h_{l+2m} &=& \delta_{m0} \label{orth-a} \\
\sum_l g_l g_{l+2m} &=& \delta_{m0} \label{orth-b}\\
\sum_l h_l g_{l+2m} &=& 0 \,, \label{orth-c}
\end{Eqnarray}
\hspace*{-0.28em}derived in the {\it Wavelet Notes}. Now using the scaling
relations in Equation (\ref{inv-ab}) gives
\begin{equation}
\phi_{j-1,m}(x) = \sum_n a_n \sum_l h_l \phi_{j-1,2n+l}(x)
                    + \sum_n b_n \sum_l g_l \phi_{j-1,2n+l}(x) \,.
\end{equation}
Taking the inner product with $\phi_{j-1,k}$ gives
\begin{Eqnarray}
\delta_{km} &=& \sum_n \sum_l a_n \,h_l \,\delta_{2n+l,k}
                  + \sum_n \sum_l b_n \,g_l \,\delta_{2n+l,k} \nonumber \\
            &=& \sum_n a_n h_{k-2n} + \sum_n b_n g_{k-2n} \,. 
\label{inveqn}
\end{Eqnarray}
\hspace*{-0.28em}Multiplying Equation (\ref{inveqn}) by $h_{m-2n^\prime}$ and
summing over $m$ gives
\begin{Eqnarray}
h_{k-2n^\prime} &=& \sum_n a_n \sum_k h_{k-2n^\prime} h_{k-2n}
                      + \sum_n b_n \sum_k h_{k-2n^\prime} g_{k-2n} \nonumber \\
                &=& \sum_n a_n \delta_{n n^\prime} \\
                &=& a_{n^\prime} \,, \nonumber
\end{Eqnarray}
\hspace*{-0.28em}where we have used the relations (\ref{orth-a}) and
(\ref{orth-c}). Multiplying Equation (\ref{inveqn}) by $g_{k-2n^\prime}$ and
summing over $k$ gives
\begin{Eqnarray}
g_{k-2n^\prime} &=& \sum_n a_n \sum_k g_{k-2n^\prime} h_{k-2n}
                      + \sum_n b_n \sum_k g_{k-2n^\prime} g_{k-2n} \nonumber \\
                &=& \sum_n b_n \delta_{n n^\prime} \\
                &=& b_{n^\prime} \,, \nonumber
\end{Eqnarray}
\hspace*{-0.28em}where we have used the relations (\ref{orth-b}) and
(\ref{orth-c}).

Given the expansion of $f(x)$ in terms of the scaling functions
\begin{equation}
f(x) = \sum_{n=0}^{2^J-1} c_{j,n} \,\phi_{j,n}(x) \,, \label{fexp}
\end{equation}
we can use Equation (\ref{inv-rel}) to write the expansion in the form
\begin{Eqnarray}
f(x) &=& \sum_{n=0}^{2^J-1} c_{j,n} \sum_m h_{n-2m} \,\phi_{j+1,m}(x)
               + \sum_{n=0}^{2^J-1} c_{j,n} \sum_m g_{n-2m} \,\psi_{j+1,m}(x)
                      \nonumber \\
     &=& \sum_{m=0}^{2^{J-1}-1} c_{j+1,m} \,\phi_{j+1,m}(x)
           + \sum_{m=0}^{2^{J-1}-1} d_{j+1,m} \,\psi_{j+1,m}(x) \,,
                 \label{fexp-2}
\end{Eqnarray}
\hspace*{-0.28em}where
\begin{Eqnarray}
c_{j+1,m} &=& \sum_{n=2m}^{2m+2p-1} h_{n-2m} \,c_{j,n} \label{c-trans} \\
\noalign{\hbox{and}}
d_{j+1,m} &=& \sum_{n=2m}^{2m+2p-1} g_{n-2m} \,c_{j,n} \,, \label{d-trans}
\end{Eqnarray}
\hspace*{-0.28em}where we use the periodic wrap-around condition $c_{j,2^J+i} =
c_{j,i}$. Equations (\ref{c-trans}) and (\ref{d-trans}) can be written as a
matrix equation, which for $J=3$ has the form
\begin{equation}
\left(\begin{array}{c}
  c_{j+1,0} \\ c_{j+1,1} \\ c_{j+1,2} \\ c_{j+1,3} \\ d_{j+1,0} \\
  d_{j+1,1} \\ d_{j+1,2} \\ d_{j+1,3}
\end{array} \right) =
\left(\begin{array}{cccccccc}
  h_0 & h_1 & h_2 & h_3 &  0  &  0  &  0  &  0  \\
   0  &  0  & h_0 & h_1 & h_2 & h_3 &  0  &  0  \\
   0  &  0  &  0  &  0  & h_0 & h_1 & h_2 & h_3 \\
  h_2 & h_3 &  0  &  0  &  0  &  0  & h_0 & h_1 \\
  g_0 & g_1 & g_2 & g_3 &  0  &  0  &  0  &  0  \\
   0  &  0  & g_0 & g_1 & g_2 & g_3 &  0  &  0  \\
   0  &  0  &  0  &  0  & g_0 & g_1 & g_2 & g_3 \\
  g_3 & g_4 &  0  &  0  &  0  &  0  & g_0 & g_1
  \end{array} \right)
\left(\begin{array}{c}
  c_{j,0} \\ c_{j,1} \\ c_{j,2} \\ c_{j,3} \\ c_{j,4} \\
  c_{j,5} \\ c_{j,6} \\ c_{j,7}
\end{array} \right) \label{mateqn-1}
\end{equation}
for the Daubechies $p=2$ wavelets. Repeated application of the filter transform
to the remaining $c_{j+1,m}$ gives
\begin{equation}
\left(\begin{array}{c}
  c_{j,0} \\ c_{j,1} \\ c_{j,2} \\ c_{j,3} \\ c_{j,4} \\
  c_{j,5} \\ c_{j,6} \\ c_{j,7}
\end{array} \right) \longrightarrow
\left(\begin{array}{c}
  c_{j+1,0} \\ c_{j+1,1} \\ c_{j+1,2} \\ c_{j+1,3} \\ d_{j+1,0} \\
  d_{j+1,1} \\ d_{j+1,2} \\ d_{j+1,3}
\end{array} \right) \longrightarrow
\left(\begin{array}{c}
  c_{j+2,0} \\ c_{j+2,1} \\ d_{j+2,0} \\ d_{j+2,1} \\ d_{j+1,0} \\
  d_{j+1,1} \\ d_{j+1,2} \\ d_{j+1,3}
\end{array} \right) \longrightarrow
\left(\begin{array}{c}
  c_{j+3,0} \\ d_{j+3,0} \\ d_{j+2,0} \\ d_{j+2,1} \\ d_{j+1,0} \\
  d_{j+1,1} \\ d_{j+1,2} \\ d_{j+1,3}
\end{array} \right) \,.
\end{equation}

The reverse transform can be obtained by substituting Equations (\ref{sc-phij})
and (\ref{sc-psij}) into Equation (\ref{fexp-2}). This gives
\begin{Eqnarray}
f(x) &=& \sum_{m=0}^{2^{J-1}-1} c_{j+1,m} \sum_{l=0}^{2k-1} h_l\,
           \phi_{j,2m+l}(x)
           + \sum_{m=0}^{2^{J-1}-1} d_{j+1,m} \sum_{l=0}^{2k-1} g_l\,
           \phi_{j,2m+l}(x) \nonumber \\
     &=& \sum_n \sum_m h_{n-2m}\,c_{j+1,m}\,\phi_{j,n}(x)
           + \sum_n \sum_m g_{n-2m}\,d_{j+1,m}\,\phi_{j,n}(x) \\
     &=& \sum_n c_{j,n}\,\phi_{j,n}(x) \,, \nonumber
\end{Eqnarray}
\hspace*{-0.28em}where
\begin{equation}
c_{j,n} = \sum_m h_{n-2m}\,c_{j+1,m} + \sum_m g_{n-2m}\,d_{j+1,m} \,.
\end{equation}
For the example used in Equation (\ref{mateqn-1}), this gives
\begin{equation}
\left(\begin{array}{c}
  c_{j,0} \\ c_{j,1} \\ c_{j,2} \\ c_{j,3} \\ c_{j,4} \\
  c_{j,5} \\ c_{j,6} \\ c_{j,7}
\end{array} \right) =
\left(\begin{array}{cccccccc}
  h_0 &  0  &  0  & h_2 & g_0 &  0  &  0  & g_2 \\
  h_1 &  0  &  0  & h_3 & g_1 &  0  &  0  & g_3 \\
  h_2 & h_0 &  0  &  0  & g_2 & g_0 &  0  &  0  \\
  h_3 & h_1 &  0  &  0  & g_3 & g_1 &  0  &  0  \\
   0  & h_2 & h_0 &  0  &  0  & g_2 & g_0 &  0  \\
   0  & h_3 & h_1 &  0  &  0  & g_3 & g_1 &  0  \\
   0  &  0  & h_2 & h_0 &  0  &  0  & g_2 & g_0 \\
   0  &  0  & h_3 & h_1 &  0  &  0  & g_3 & g_1
  \end{array} \right)
\left(\begin{array}{c}
  c_{j+1,0} \\ c_{j+1,1} \\ c_{j+1,2} \\ c_{j+1,3} \\ d_{j+1,0} \\
  d_{j+1,1} \\ d_{j+1,2} \\ d_{j+1,3}
\end{array} \right) \,, \label{mateqn-2}
\end{equation}
which is the transpose of the matrix in Equation (\ref{mateqn-1}).

\section{Wavelet Transfrom}

The wavelet transform is by its nature local at each level and
therefore admits an implementation in which the data to be transformed
can be placed in a buffer instead of storing the entire data set at
once.  This significantly reduces the amount of storage space required
for applications involving compression.  

In the one-dimensional case,
the $J$-level wavelet transform can be computed by buffering $O(J)$
nonessential elements or the full transform can be computed buffering
$O(\log(N))$ elements.  The standard form for the two-dimensional
transform of an $N \times N$ matrix can be performed by buffering only
$O(N\log(N))$ elements.  In general, a $D$-dimensional wavelet transform
can be computed by only storing $O(N^{D-1}\log(N))$ elements.  This
buffered wavelet transform can be applied to any type of data that can
be input or computed in series.  Some notable examples include the
compression of time-series data and applications to solutions of
integral equations.  Below, we will explain the exact implementation of
the transform including the buffer and the extension of the method to
two dimensions.  The extension to arbitrary dimension is
straightforward from the two dimensional case.  First, we will layout
the terminology that will be used throughout.  we adopt the filter
viewpoint since it makes the explanation of the buffering procedure
more clear.  But this is equivalent to the linear algebra viewpoint
and we will attempt to explain the procedure in this language as well.

From the filter viewpoint, the wavelet transform is a convolution of
the data set and two vectors $h$ and $g$ followed by a decimation.  This
is equivalent to a convolution that proceeds by steps of two instead
of one.  For the Daubechies family of wavelets, both of these filters
have a length $L = 2K-1$.  The convolution with $h$ produces what is
called the father set and the convolution with $g$ produces the mother
set.  We denote the data set to be transformed as A and use brackets to
denote subscripts.  In all contexts below, the one-dimensional length
of the data set will be $N = 2^n$ where $n$ is an integer.  That is the
data runs from $A[0]$ to $A[N-1]$.  The typical procedure to deal with a
finite data set is to periodize the data over the boundary such that
$A[N] = A[0], A[N+1]=A[1], \cdots A[N+L-3]=A[L-3]$.  

Now, if we write out
the first level of the transform as a matrix we can see that it is
banded with a bandwidth L corresponding to the convolution operation.
The second level of the transform is identical to the first except
that it only acts on the father set of data, i.e. the transform on the
mothers is the identity.  This corresponds to the fact that all of the
information about coarser levels is contained in the father functions.
The mother functions form an orthogonal subspace to the fathers and
mothers on all higher levels.  Using this knowledge we can immediately
perform a thresholding procedure on the mother set without affecting
the rest of the data in any way.  The father set can simultaneously be
transformed and the resulting mothers thresholded as well.  Iterating
this procedure on all the relevant levels forms the basis for the
buffered wavelet transform.  Of course, one must have an a priori
thresholding scheme to accomplish this.  The simplest such example is
an absolute threshold.  In this scheme, one chooses an epsilon and all
elements with a magnitude less than this epsilon.  Other more
sophisticated thresholding procedures exist as well, such as
procedures based on the level on the transform.  The important fact is
that one cannot have a procedure that depends in any way on the final
transformed data set.  Examples of such procedures would be based on
the relative size of the transformed elements or a threshold that
keeps a certain number/percentage of the final coefficients.  In many
applications, the absolute thresholding is an acceptable method.  

Now,
we will explain the detailed implementation of the one-dimensional
transform.  One begins by computing the elements $A[0]$ to $A[L-3]$ of the
data set.  As noted above, these elements are necessary for the
periodic boundary conditions and form a boundary buffer that must be
saved until the end of the calculation.  Now, elements can be added to
a moving buffer of length $L$ that constitutes the heart of the
procedure.  After the elements $A[L-2]$ and $A[L-1]$ are computed and
placed in the moving buffer, one can begin transforming the data set.
Convolving this data set, including the boundary terms, with h
produces the first member of the father set $F[0]$ and convolving with g
produces the first member of the mother set $M[0]$.  As described
previously, this mother element can be immediately thresholded and
placed in the final output vector.  The father element is considered
the beginning of a new data set to be transformed and is placed in the
boundary buffer corresponding to the next level of the transform.  One
then proceeds to compute two more elements and convolve $A[2]$ to $A[L+1]$
with $h$ and $g$.  This produces $F[1]$ and $M[1]$, which are treated the same
way as before.  We continue in this manner, calculating more elements
and convolving, until we have computed the element $A[2L-3]$.  The
moving buffer is now full and we have reached the interior of the data
set.  When we compute the next element of the data set we can discard
the last element of the moving buffer and shift all the elements of
the buffer one place.  The new element is then appended to the moving
buffer.  Discarding the last element is justified by the fact that all
the information in that element is represented by the corresponding
father and mother data sets due to the equivalence of the subspaces.
The name moving buffer is clear since this buffer can be viewed as
scanning the interior of the data set by moving over it.  This process
continues, the shifting and convolving, until the end of the data set
is reached.  When the end of the data set is reached we simply make
the data set periodic using the boundary buffer.  This process is
simultaneously carried out at each level.  Now counting the elements
in each buffer we see that in each level we must store $L-2$ elements in
the boundary buffer and $L$ elements in the moving buffer.  So, for $J$
levels we must store $J(2L-2)$ elements.  This gives us our size of
$O(J)$ where the coefficient depends on the length $L$ of the wavelet
filter as is common with most wavelet algorithms.  

In many wavelet
applications, the data vector to be transformed will be of length
$N=2^n$ and a wavelet transform of level $J=n$ will be computed.  In this
case, the number of elements stored in the buffers will be of
$O(\log(N))$.  A minor point to note is that for wavelet filters of
length $> 2$ the last few levels will not be filled completely.  As a
programming point one can either fill the buffers periodically or just
periodize the convolution.  Both procedures are equivalent and
consistent with the periodic wavelet transform.  Also note that the
number of operations has been increased by the shift operation, but is
still of $O(N)$ which is the case for the standard wavelet transform.
The standard procedure to perform the wavelet transform on a two
dimensional data set is to first transform the rows of the matrix and
then transform the columns.  Alternatively, one could transform the
columns and then the rows.  Both are equivalent as can be seen by writing out
the transform as a matrix multiply and noting the associativity of
matrix multiplication.  

To perform the buffered wavelet transform on a
two-dimensional data set we calculate the data column by column.  Each
row has a separate set of buffers associated with it.  We can view
this as a strip that scans the matrix in much the same way as the
moving buffer did in the one-dimensional case.  Each of these buffers
behaves in exactly the same way as the one-dimensional case, except
the output is handled differently.  The first output from the buffer
associated with row one is placed in two vertical buffers.  $FB[0]$ and
$MB[0]$ the $B$ stands for blank since these are internal buffers that
have no outside significance.  Both of these outputs must be saved
because they contain information about the columns of the matrix.  Row
two then produces $FB[1]$ and $MB[1]$, and so on continuing down the rows.
The transform procedure is applied to the vertical buffers, which
produce output $FF$, $FM$, $MF$, and $MM$.  The output $MM$ can be thresholded
immediately.  The output $FM$ is placed in an array of row buffers of
height $N/2$ that transform the rows and filter immediately the $MB$'s
produced. The $MF$ output is placed in another vertical buffer where the
traditional one-dimensional transform procedure is enacted.  The $FF$
output is placed in an array of row buffers identical to the original
configuration, except that it is only $N/2$ tall.  The same procedure is
enacted on this data set that is half as small.  Now this proceeds
across the matrix in a similar manner as the one-dimensional case,
except that the vertical buffers can be completely purged as the next
column is reached.  To count the number of elements that are necessary
we can ignore the vertical buffers, which are subdominant.  At the
first level we note that there are $N*(2L-2)$ elements in the row buffer
and $(N/2)*(2L-2)$ in the $FF$ output and $FM$ output.  Hereafter we will
drop the $2L-2$ to simplify the counting since we are just looking for
order.  So we have $N$ and $N/2$ and $N/2$.  The $FM$ output will proceed like
the one-dimensional case.  Therefore it will produce $(N/2)\log(N)$
elements.  The $FF$ out put will produce $N/4 FF$ and $N/4 FM$.  So we can
see that the total number of elements will be
$N(1+\log (N))\sum_{j=0}^J{1\over 2}^j$.  The sum is a simple geometric sum
that in the limit that $J$ goes to infinity is bounded by $2$.  So the
final tally of the necessary elements is $O(N\log (N))$.  The
generalization to $D$ dimensions is straightforward.  One begins with a
data structure of dimensions $(N^{D-1})(2L-2)$.  One then performs a
transform to produce two $N^{D-2}$ data structures.  One performs a
transform on these two structures to produce four $N^{D-3}$ structures.
This process continues until the final transform where one has a
single dimension.  This transform is enacted and the $M^{D}$ elements
are filtered.  Appropriate lower dimensional transforms are applied to
the mixed output, $MMMFF\cdots MF, FFFFFF \cdots M,$ etc.  The process is
repeated for the $FFF\cdots FFF$ data set.  In higher dimensions the
algorithm becomes more complicated, but the idea is the same.  And the
leading order number of elements that need to be saved is
$O(N^{D-1}\log (N))$.

\section{Appendix I: Continuous Wavelets} 

We begin by considering the continuous wavelet transform.  The
continuous wavelet transform is an alternate representation of a
function, like a Fourier transform.  Both continuous and discrete
wavelets are built from a single function called a {\bf mother
function}.  The notation, $\psi(x)$, is used to denote the mother
function of a wavelet.

Wavelets are built from translations and scale transformations of 
the mother function. Translations and scale transformations of $\psi(x)$
are defined by:
\beq
\psi_{t,s} (x) := \vert s \vert^{-p}\psi({x-t \over s}).
\eeq

The factor $p$ is a parameter.  The functions $\psi_{t,s} (x)$ are the
wavelets associated with the mother function $\psi(x)$.  The wavelet
$\psi_{t,s} (x)$ has two continuous parameters.  We investigate
conditions on the mother function that allow one to expand any
function in terms of wavelets.

To choose the parameter $p$ note that
\[
\int_{-\infty}^{\infty} \left\vert \vert s \vert^{-p}\psi({x-t \over s})
\right\vert^q dx =
\]
\beq
\vert s \vert^{1 - qp}  \int_{-\infty}^{\infty} \vert \psi(u) \vert^q du . 
\eeq
It follows that if $p=1/q$ the $L^q$-norm of $\psi$  
\beq
\Vert \psi \Vert_q := \left( \int_{-\infty}^{\infty} \vert \psi(u) \vert^q du
\right)^{1/q}  
\eeq
is preserved under scale transformations.  Thus for $p=1/q$:

\beq
\Vert \psi \Vert_q = \Vert \psi_{s,t} \Vert_q \qquad \mbox{for all} 
\qquad s,t .
\eeq

The {\bf continuous wavelet transform} of $f$ is defined by taking the 
scalar product of $f$ with the wavelet $\psi_{s,t}$:
\beq
\hat{f}(s,t) :=  
\int_{-\infty}^{\infty} \psi^*_{s,t}(x) f(x) dx = (\psi_{s,t},f) 
\eeq
where asterik $'*'$ indicates the complex conjugate for a
complex mother function.  In what follows a $\hat{f}$ is used to
indicate the wavelet transform of the function $f$.

Parseval's identity for the Fourier transform implies that the wavelet 
transform can be expressed in terms of the original 
function and the mother function 
or alternatively in terms of their Fourier transforms:
\beq
\hat{f}(s,t)= (\psi_{s,t},f) = (\tilde{\psi}_{s,t},\tilde{f})
\label{eq:AA}
\eeq
where the $\sim$ indicates the Fourier transform defined by:
\beq
\tilde{\psi}_{s,t}(k) = {1 \over \sqrt{2 \pi}} \int_{-\infty}^{\infty}
e^{-ikx} \psi_{s,t} (x) dx
\eeq
\beq
\tilde{f}(k) = {1 \over \sqrt{2 \pi}} \int_{-\infty}^{\infty}
e^{-ikx} f(x) dx.
\eeq

Note that Parseval's identity states 
$(f,f) = (\tilde{f},\tilde{f})$. 
Using this with $f=g+h$ and $f=g+ih$ gives
\beq
(g,g) + (h,h) + (g,h) + (h,g) =(\tilde{g},\tilde{g}) + 
(\tilde{h},\tilde{h}) + (\tilde{g},\tilde{h}) + (\tilde{h},\tilde{g}) 
\label{eq:AB}
\eeq
and
\beq
(g,g) + (h,h) + i(g,h) -i(h,g) =(\tilde{g},\tilde{g}) + 
(\tilde{h},\tilde{h}) +i (\tilde{g},\tilde{h}) -i (\tilde{h},\tilde{g}) 
\label{eq:AC}
\eeq
which, using the identities $(g,g)=(\tilde{g},\tilde{g})$ and 
$(h,h)=(\tilde{h},\tilde{h})$, gives the solution to (\ref{eq:AB})
and (\ref{eq:AC}):
\beq
(g,h)=(\tilde{g},\tilde{h})
\eeq
which is the form of Parseval's identity used in (\ref{eq:AA}).

The Fourier transform of the wavelet 
$\psi_{s,t}(x)$ can be expressed in terms of the 
Fourier transform of the mother function:

\[
\tilde{\psi}_{s,t}(k) := {1 \over \sqrt{2 \pi} }\int_{-\infty}^{\infty}
 e^{-i kx} \vert s \vert^{-p} \psi({x-t \over s}) dx=
\]
\[
{1 \over \sqrt{2 \pi} }\int_{-\infty}^{\infty}
e^{- i ksu} e^{- i kt} \vert s \vert^{-p+1} \psi(u) du=
\]
\beq
\vert s \vert^{1-p} e^{- i kt} \tilde{\psi} (sk).
\eeq

The inner product of the Fourier transforms gives
\[
\hat{f}(s,t) = (\tilde{\psi}_{s,t},\tilde{f})=
\]
\[
\int_{-\infty}^{\infty} \tilde{\psi}_{s,t}^* (k)
\tilde{f}(k) dk 
\]
\beq
\int_{-\infty}^{\infty} \vert s\vert^{1-p} e^{ i kt} \tilde{\psi}^* (sk)
\tilde{f}(k) dk . 
\label{eq:FA}
\eeq
Multiplying both sides of (\ref{eq:FA}) by $e^{-i k't}$ and 
integrating over $t$ gives
\[
{1 \over 2 \pi} \int_{-\infty}^{\infty} e^{- i k't} 
(\tilde{\psi}_{s,t},\tilde{f})dt =
\]
\beq
\vert s \vert^{1-p} \tilde{\psi}^* (sk')
\tilde{f}(k'),
\label{eq:FB}
\eeq
where the representation of the delta function:
\beq
{1 \over 2 \pi} \int_{-\infty}^{\infty} e^{- i (k'-k)t}dt = \delta (k'-k).
\eeq
was used to get (\ref{eq:FB}).

The right-hand side of (\ref{eq:FB}) is a product of the Fourier 
transform of the original function with another function.  
We can't divide by the function $\tilde{\psi}^* (sk')$ 
because it might be zero for some values of $k'$.  Instead, the trick is to
eliminate it using the variable $s$.

Multiply both sides of this equation by $\tilde{\psi}(sk')$ and a 
yet to be determined weight function $w(s)$ and integrate over $s$.
This gives

\[
{1 \over 2 \pi} \int_{-\infty}^\infty w(s) ds \int_{-\infty}^{\infty} dt 
e^{- i k't} \tilde{\psi}(sk') \hat{f}(s,t) =
\]
\beq
\tilde{f}(k') \int_{-\infty}^\infty w(s) ds
\vert s \vert^{1-p} \tilde{\psi}^* (sk') \tilde{\psi}(sk') =
\tilde{f}(k') Y(k') 
\label{eq:FC}
\eeq 
where
\beq
Y(k')= \int_{-\infty}^\infty ds w(s)  \vert s \vert^{1-p} 
\vert \tilde{\psi}(sk')\vert^2 .
\eeq

In order to be able to extract the Fourier transform of the original function,
it is sufficient that $Y(k')$ satisfies $0 < A \leq Y(k') \leq B < \infty$  
for some numbers $A$ and $B$.  In this case

\beq
\tilde{f} (k) = { 1 \over 2 \pi Y(k)}   
\int_0^\infty w(s) ds \int_{-\infty}^{\infty} dt 
e^{- i kt} \tilde{\psi}(sk) \hat{f}(s,t) .
\eeq

It is convenient to rewrite this in terms of the wavelet basis:
\beq
\tilde{f} (k) = { 1 \over 2 \pi Y(k)}   
\int_{- \infty}^\infty w(s) \vert s \vert^{p-1} ds \int_{-\infty}^{\infty} dt 
\tilde{\psi}_{s,t}(k) \hat{f}(s,t). 
\eeq

We define the {\bf dual wavelet} by
\beq
\tilde{\psi}^{s,t}(k) =
{ 1 \over 2 \pi Y(k)}\tilde{\psi}_{s,t}(k).
\label{eq:AD}
\eeq
The dual wavelet is distinguished from the ordinary wavelet by having 
the parameters $s,t$ appearing as superscripts rather than subscripts.

The inversion formula can be expressed in terms of the dual wavelet by
\beq
\tilde{f} (k) =    
\int_{-\infty}^\infty w(s) \vert s \vert^{p-1} ds \int_{-\infty}^{\infty} dt 
\tilde{\psi}^{s,t}(k) \hat{f}(s,t) .
\eeq

In order to recover the original function, take the inverse Fourier 
transform of this expressions:
\[
f(x) =  {1 \over \sqrt{2 \pi}} \int^{\infty}_{-\infty} dk
e^{ikx}\tilde{f}(k) =
\]
\beq
\int_{-\infty}^\infty w(s) \vert s\vert^{p-1} ds \int_{-\infty}^{\infty} dt 
\psi^{s,t}(x) \hat{f}(s,t) 
\eeq
where 
\beq
\psi^{s,t}(x) = {1 \over \sqrt{2 \pi}} \int^{\infty}_{-\infty} dk
e^{ikx} \tilde{\psi}^{s,t}(k).
\label{eq:AE}
\eeq
In general this is a tedious procedure because the dual wavelet $\psi^{s,t}(x)$
must be computed using (\ref{eq:AD}) and (\ref{eq:AE})
for each value of $s$ and $t$.  If the dual wavelet also
had a mother function, then it would only be necessary to Fourier transform 
the ``dual mother'' and then all of the other Fourier transforms could be 
expressed in terms of the transform of the ``dual mother''. 

The first step in constructing a ``dual mother'' is to investigate the 
structure of the dual wavelets in $x$-space:

\[
\psi^{s,t}(x) = {1 \over \sqrt{2 \pi}} \int^{\infty}_{-\infty} dk
e^{ikx} \tilde{\psi}^{s,t}(k) = 
\]
\[
{1 \over \sqrt{2 \pi}} \int^{\infty}_{-\infty} dk
e^{ikx}  { 1 \over 2 \pi Y(k)}
\vert s \vert^{1-p} e^{-i kt} \tilde{\psi} (sk) =
\]
\[
\psi^{s,0}(x-t) 
\]
where
\[
\psi^{s,0}(x) =
{1 \over \sqrt{2 \pi}} \int^{\infty}_{-\infty} dk
e^{ikx}  { 1 \over 2 \pi Y(k)}
\vert s \vert^{1-p} \tilde{\psi} (sk).
\]
This shows for a single scale the dual wavelet and its translation can 
be expressed in terms of a single function.  This is not necessarily 
true for the dual wavelet and the scaled quantity. 
\[
\psi^{s,0}(x) =
{1 \over \sqrt{2 \pi}} \int^{\infty}_{-\infty} dk
e^{ikx}  { 1 \over 2 \pi Y(k)}
\vert s\vert^{1-p} \tilde{\psi} (sk) 
\]
\[
=
{1 \over \sqrt{2 \pi}} \int^{\infty}_{-\infty} du
e^{iu {x \over s} }  { 1 \over 2 \pi Y(u/s)}
\vert s\vert^{-p} \tilde{\psi} (u).
\]
This fails to be a rescaling of a single function because of 
the $s$ dependence 
in the quantity $Y(u)$.  It follows that if {\it a weight function
$w(s)$} is chosen so $Y(u/s)=Y$ is constant,  the dual wavelet will
satisfy
\beq
\psi^{s,0}(x) =
{1 \over \sqrt{2 \pi}} \int^{\infty}_{-\infty} du
e^{iu {x \over s} }  { 1 \over 2 \pi Y}
\vert s\vert^{-p} \tilde{\psi} (u) =
\vert s\vert^{-p} \psi^{1,0}(x/s). 
\eeq
In this case $Y(u)$ is a constant which we denote by $Y$.  
The function $\psi^{1,0}(x)$ serves as the dual mother wavelet.

To determine $w(s)$ note
that 
\[ 
Y(sk) = \int_{-\infty}^{\infty} dt w(t) \vert t\vert^{1-p} \vert \tilde{\psi} 
(tsk)\vert^2.
\]
Let $t' = st$ to get
\[ 
Y(sk) = \int_{-\infty}^{\infty} dt w(t) \vert t\vert^{1-p} \vert \tilde{\psi} (tsk)\vert^2 =
\]
\[
\vert s\vert^{p-2} \int_{-\infty}^{\infty} dt' 
w(t'/s) \vert t\vert^{\prime 1-p} \vert \tilde{\psi} (t'k)\vert^2 .
\]
This will equal $Y(k)$ provided
\[
w(t') = \vert s\vert^{p-2} w(t'/s) \qquad \mbox{or} \qquad w(s) = \vert s\vert^{p-2}w(1) .
\]

With this choice 
\[ 
Y(k) = Y= w(1) \int_{-\infty}^{\infty} {dt \over \vert t\vert} \vert \tilde{\psi} 
(t)\vert^2  .
\]
Assuming this choice of weight the admissibility condition becomes
\[
0 < A \leq Y  \leq B < \infty .
\]

Having computed the constant $Y$ it is now possible to write down
an explicit expression for the dual mother wavelet:
\[
\psi^{s,0}(x-t) =
\]
\[
{\vert s\vert^{-p} \over \sqrt{2 \pi}}  \int^{\infty}_{-\infty} du
e^{iu {(x-t) \over s} }  { 1 \over 2 \pi Y}
\tilde{\psi} (u)
 \]
Letting $k=u/s$ 
\[
{1 \over \sqrt{2 \pi}} \int^{\infty}_{-\infty} dk
{ 1 \over 2 \pi Y}
\vert s\vert^{1-p} e^{ik(x-t) } \tilde{\psi} (ks)
\]
\[
{1 \over \sqrt{2 \pi}} \int^{\infty}_{-\infty} dk
{ 1 \over 2 \pi Y}
e^{ikx } \tilde{\psi}_{s,t} (k) .
\]
This has the form
\beq
\psi^{s,t}(x) = 
{1 \over 2 \pi} {1 \over Y}\psi_{s,t}(x).
\eeq

Thus the inversion procedure can be summarized by the formulas:
\beq
f(x) =
\int_{-\infty}^\infty \vert s\vert^{2p-3} ds \int_{-\infty}^{\infty} dt 
\psi^{s,t}(x) \hat{f}(s,t) 
\eeq
\beq
Y =\int_{-\infty}^{\infty} {dt \over \vert t\vert} \vert \tilde{\psi} 
(t)\vert^2  
\eeq
\beq
\psi^{s,t}(x) = {\psi_{s,t}(x) \over 2 \pi Y}
\eeq
\beq
\psi_{s,t} =\vert s \vert^{-p} \psi({x-t \over s}).  
\eeq

The mother function must satisfy $0< Y< \infty$.  
This requires that the Fourier transform of the mother function 
vanishes at the origin.  This is equivalent to saying that
the integral of the mother function is zero.

Using the representation for the wavelet transform gives a representation 
of a delta function:

\[
\delta (x-y) =
\]
\[
\int_{-\infty}^\infty \vert s\vert^{2p-3} ds \int_{-\infty}^{\infty} dt 
\psi^{s,t}(x) \psi^*_{s,t} (y)  =
\]
\[
{1 \over 2 \pi Y} \int_{-\infty}^\infty 
\vert s\vert^{2p-3} ds \int_{-\infty}^{\infty} dt 
\psi_{s,t}(x) \psi^*_{s,t} (y) .
\]
We can also use this representation of the delta function 
to formulate a Parseval's identity for continuous 
wavelets
\beq
(f,f) = {1 \over 2 \pi Y} \int_{-\infty}^\infty 
\vert s\vert^{2p-3} ds \int_{-\infty}^{\infty} dt 
\vert \hat{f} (s,t) \vert^2 .
\eeq

Consider the example of the {\bf Mexican hat} wavelet.  The mother 
function is
\[
\psi(x) = {1 \over \sqrt{2 \pi} } (x^2 -1) e^{-x^2/2}.
\]

To work with the Mexican hat mother function it is useful to use
properties of Gaussian integrals:
\[
\int_{-\infty}^{\infty} e^{-ax^2 + bx +c} dx =
\]
\[
\int_{-\infty}^{\infty} e^{-a(x- {b\over 2a})^2 + {b^2\over 4a} +c} dx.
\]
Change variables to $y = \sqrt{a}(x- {b\over 2a})$ to obtain:
\[
{ e^{{b^2\over 4a} +c} \over \sqrt{a}} \int_{-\infty}^{\infty} 
e^{-y^2} dy =
\]
\[
\sqrt{{\pi \over a}} e^{{b^2\over 4a} +c}. 
\]

This can be used to compute the 
Fourier transform of the Mexican hat mother function: 
\[
\tilde{\psi}(k) = {1 \over \sqrt{2 \pi}}\int_{-\infty}^{\infty}
e^{-ikx} \psi(x) dx =
\]
\[
{1 \over {2 \pi}}\int_{-\infty}^{\infty}
(x^2 -1) e^{-x^2/2 -ikx} dx .
\]
To do the integral insert a parameter $a$ which will be set to 1 at the 
end of the calculation:
\[
( - 2{d \over da} -1 ) {1 \over {2 \pi}}\int_{-\infty}^{\infty}
e^{-x^2a /2 -ikx} dx  =
\]
\[
( - 2{d \over da} -1 ){1 \over {2 \pi}}
\sqrt{{2\pi \over a}} e^{-{k^2\over 2a}} =
\]
\[
( {1 \over a} - {k^2 \over a^2} -1)   
\sqrt{{1 \over 2 \pi a}} e^{-{k^2\over 2a}}.
\]
In the limit that $a \to 1$ this becomes
\[
- \sqrt{{1 \over 2 \pi }}k^2  e^{-{k^2\over 2}}.
\] 

Using this expression it is possible to calculate the coefficient $Y$
\[
Y = \int_{-\infty}^{\infty} {dk \over \vert k \vert} \vert \tilde{\psi} (k)
\vert^2 =
\]
\[
\int_{-\infty}^{\infty} {dk \over \vert k \vert} \vert \tilde{\psi} (k)
\vert^2 =
\]
\[
{1 \over 2 \pi } \int_{-\infty}^{\infty} \vert k \vert^3 dk   
e^{-{k^2}} =
\]
\[
{1 \over \pi } \int_{0}^{\infty} k^3 dk   
e^{-{k^2}}.
\]
Inserting a parameter $a$ which will eventually be set to 1 gives
\[
Y={1 \over 2\pi }(- {d \over d a}) \int_{0}^{\infty} 2k dk   
e^{-ak^2} =
\]
\[
{1 \over 2\pi }(- {d \over d a}){1 \over a}  \int_{0}^{\infty} dv   
e^{-v} =
\]
\[
{1 \over 2\pi }.
\]
This satisfies the essential inequality $0<Y < \infty$ which ensures
the admissibility of the Mexican hat mother function.

The expression for the wavelet transform and its 
inverse can be written as:

\[
\hat{f}(s,t) = \vert s \vert^{-p}\int_{-\infty}^{\infty}dx 
{1 \over \sqrt{2 \pi} } (({x-t \over s})^2 -1) e^{-({x-t \over s})^2/2}
f(x) = 
\]
\[
\vert s \vert^{1-p}\int_{-\infty}^{\infty}du 
{1 \over \sqrt{2 \pi} } (u^2 -1) e^{-u^2/2}
f(su+t).
\]
where $x=su+t$

The inverse is formally given by 

\[
f(x) = \int_{-\infty}^{\infty} \vert s \vert^{2p-3} ds \int_{-\infty}^{\infty} 
dt {\psi_{st}(x) \over 2 \pi Y} \hat{f}(s,t) =
\]
\[
\int_{-\infty}^{\infty} \vert s \vert^{2p-3} ds \int_{-\infty}^{\infty} 
dt  {1 \over \sqrt{2 \pi} } \vert s \vert^{-p} 
(({x-t \over s})^2 -1) e^{-({x-t \over s})^2/2} \hat{f}(s,t) =
\]
\[
{1 \over \sqrt{2 \pi} } 
\int_{-\infty}^{\infty} \vert s \vert^{p-3} ds \int_{-\infty}^{\infty} 
dt   
(({x-t \over s})^2 -1) e^{-({x-t \over s})^2/2} \hat{f}(s,t) =
\]
\[
{1 \over \sqrt{2 \pi} } 
\int_{-\infty}^{\infty} \vert s \vert^{p-3} ds \int_{-\infty}^{\infty} 
du   
(u^2 -1) e^{-u^2/2} \hat{f}(s,su+x)
\]
where $t=su+x$.

Initially we were concerned because we were representing an arbitrary 
function by a linear superposition of functions that all have zero integral.
We could not understand how wavelets could be used to represent a function
with non-zero integral.

We tested this by computing the wavelet transform and its inverse 
for a Gaussian function with the Mexican hat wavelet.  The original 
Gaussian function was recovered.  

The resolution of this paradox has to do with the difference between $L^1$
and $L^2$ convergence.  The wavelet transform has a vanishing $L^1$ norm,
but the $L^2$ norm is non-zero.

\section{Appendix II - Spline Wavelets}
\bigskip
We use the convention which defines the Fourier transform of a function $f(x)$
as
\begin{equation}
F(k) = \int_{-\infty}^\infty e^{-ikx} f(x)\,dx \,,
\end{equation}
and the inverse transform by
\begin{equation}
f(x) = \frac{1}{2\pi}\int_{-\infty}^\infty F(k)e^{ikx}\,dk \,.
\end{equation}
For this convention, Parseval relation is
\begin{equation}
\int_{-\infty}^\infty f^\ast(x)g(x)\,dx = \frac{1}{2\pi}\int_{-\infty}^\infty
     F^\ast(k) G(k)\,dk \,.
\end{equation}

The cardinal B-splines, $N_m(x)$, are defined by first defining
\begin{equation}
N_1(x) = \cases{0, &if $x<0$ \cr
                1, &if $0<x<1$ \cr
                0, &otherwise. \cr}
\end{equation}
Then for $m \ge 2$, $N_m(x)$ is defined recursively by the convolution integral
\begin{eqnarray}
N_m(x) &=& \int_{-\infty}^\infty N_{m-1}(x-t) N_1(t)\,dt \nonumber\\
       &=& \int_0^1 N_{m-1}(x-t)\,dt \,.
\end{eqnarray}
Since $N_m(x)$ is defined by a convolution integral, the Fourier transform will
be defined by a product. To show this, we evaluate the Fourier transform
\begin{eqnarray}
\tilde{N}_m(k) &=& \int_{-\infty}^\infty e^{-ikx} N_m(x)\,dx \nonumber\\
             &=& \int_{-\infty}^\infty dx\,e^{-ikx}\int_{-\infty}^\infty
                      N_{m-1}(x-t) N_1(t)\,dt \nonumber\\
             &=& \int_{-\infty}^\infty dt\,e^{-ikt} N_1(t)\int_{-\infty}^\infty
                      e^{-ik(x-t)} N_{m-1}(x-t)\,dx \,.
\end{eqnarray}
Now setting $x-t=y$ in the second integral, we find
\begin{eqnarray}
\tilde{N}_m(k) &=& \int_{-\infty}^\infty dt\,e^{-ikt} N_1(t) 
               \int_{-\infty}^\infty e^{-iky} N_{m-1}(y)\, dy \nonumber\\
           &=& \tilde{N}_{m-1}(k) \int_{-\infty}^\infty e^{-ikt} N_1(t) \,dt 
               \nonumber\\
           &=& \tilde{N}_{m-1}(k)\tilde{N}_1(k) \nonumber\\
           &=& \left[\tilde{N}_1(k)\right]^m \,,
\end{eqnarray}
where
\begin{eqnarray}
\tilde{N}_1(k) &=& \int_{-\infty}^\infty e^{-ikx} N_1(x)\,dx \nonumber\\
               &=& \int_0^1 e^{-ikx}\,dx \nonumber\\
               &=& e^{-ik/2} \frac{\sin(k/2)}{k/2} \,.
\end{eqnarray}

For this example we use the quadratic spline shifted to the left by one unit
\begin{eqnarray}
\tilde{B}(k) &=& e^{ik}\tilde{N}_3(k) \nonumber\\
             &=& e^{-ik/2}\left[\frac{\sin(k/2)}{k/2}\right]^3 \,. \label{qbk}
\end{eqnarray}
Now evaluating the inverse transform using Maple, we get
\begin{equation}
B(x) = \cases{0, &if $x < -1$ \cr
                \frac{1}{2}(x+1)^2, &if $-1 \le x < 0$ \s\cr
                \frac{3}{4}-(x-\frac{1}{2})^2, &if $0 \le x < 1$ \s\cr
                \frac{1}{2}(x-2)^2, &if $1 \le x < 2$ \s\cr
                0, &otherwise. \s\cr}
\end{equation}
The splines are not orthogonal; however, we can use them to construct a
scaling function $\phi(x)$ which has the orthonormality property
\begin{equation}
\int_{-\infty}^\infty \phi^\ast(x-l) \phi(x-m)\,dx = \delta_{lm}\,. 
                      \label{ortho}
\end{equation}
To do this we follow the procedure given in the books by Chui\footnote{Charles
K. Chui, \textit{An Introduction to Wavelets}, Academic Press, 1992} and
Daubechies\footnote{Ingrid Daubechies, \textit{Ten Lectures on Wavelets},
SIAM, 1992}. Note that this a general procedure; we are using the spline as a
convenient example. The method gives an expression for the Fourier transform,
$\tilde{\phi}(k)$, of $\phi(x)$. The Fourier transform of $\phi(x-l)$ is given
by
\begin{eqnarray}
\tilde{\phi}_l(k) &=& \int_{-\infty}^\infty e^{-ikx} \phi(x-l)\,dx \nonumber\\
                  &=& e^{-ikl}\int_{-\infty}^\infty e^{-ik(x-l)} \phi(x-l) \,dx
                        \nonumber\\
                  &=& e^{-ikl}\tilde{\phi}(k) \,.
\end{eqnarray}
Now we show that if
\begin{equation}
\frac{1}{2\pi}\int_{-\infty}^\infty e^{ikm} \left|\tilde{\phi}(k)\right|^2\,dk
      = \delta_{m,0} \,,
\end{equation}
then the functions are orthogonal. To show this, we use the Parseval relation
\begin{eqnarray}
\int_{-\infty}^\infty \phi^\ast(x-l) \phi(x-m)\,dx 
     &=& \frac{1}{2\pi}\int_{-\infty}^\infty \tilde{\phi_l}^\ast(k)
            \tilde{\phi_m}(k)\,dk \nonumber\\
     &=& \frac{1}{2\pi}\int_{-\infty}^\infty e^{ikl}\tilde{\phi}^\ast(k)
            e^{-ikm}\tilde{\phi}(k) \,dk \nonumber\\
     &=& \frac{1}{2\pi}\int_{-\infty}^\infty e^{ik(l-m)}\left|\tilde{\phi}(k)
            \right|^2\,dk \nonumber\\
     &=& \delta_{l-m,0} \nonumber\\
     &=& \delta_{lm}\,.
\end{eqnarray}

Finally, we show that if we can find a $\tilde{\phi}(k)$ such that
\begin{equation}
\sum_{l=-\infty}^\infty \left|\tilde{\phi}(k+2\pi l)\right|^2 = 1 \,, 
               \label{grel}
\end{equation}
then the functions are orthonormal. The infinite sum in Equation (\ref{grel})
is periodic in $k$ with a period of $2\pi$; thus it has the Fourier series
expansion
\begin{equation}
\sum_{l=-\infty}^\infty \left|\tilde{\phi}(k+2\pi l)\right|^2
          = \sum_{j=-\infty}^\infty c_j e^{ikj}\,
\end{equation}
where the expansion coefficients are give by
\begin{eqnarray}
c_j &=& \frac{1}{2\pi} \int_0^{2\pi} e^{-ijk} \sum_{l=-\infty}^\infty 
          \left|\tilde{\phi}(k+2\pi l)\right|^2\,dk \nonumber\\
    &=& \frac{1}{2\pi} \sum_{l=-\infty}^\infty \int_0^{2\pi} e^{-ikj}
          \left|\tilde{\phi}(k+2\pi l)\right|^2 \,dk \nonumber\\
    &=& \frac{1}{2\pi} \sum_{l=-\infty}^\infty \int_{2\pi l}^{2\pi(l+1)}
          e^{-i(k-2\pi l)j}\left|\tilde{\phi}(k)\right|^2\,dk \nonumber\\
    &=& \frac{1}{2\pi} \int_{-\infty}^\infty e^{-ikj}\left|\tilde{\phi}(k)
          \right|^2\,dk \,. \label{coef}
\end{eqnarray}
Since the sum in Equation (\ref{grel}) is equal to one, $c_j=\delta_{j,0}$,
and one finds
\begin{equation}
\frac{1}{2\pi} \int_{-\infty}^\infty e^{-ikj}\left|\tilde{\phi}(k) \right|^2
         \,dk = \delta_{j,0}\,.
\end{equation}
Thus, the functions are orthonormal.

Now given a function, $B(x)$, we construct a scaling function by taking its
Fourier transform and defining
\begin{equation}
\tilde{\phi}(k) = \frac{\tilde{B}(k)}{\displaystyle \left[
  \sum_{l=-\infty}^\infty \left| \tilde{B}(k+2\pi l)\right|^2
  \right]^\frac{1}{2}} \,. \label{scfuna}
\end{equation}
This function satisfies Equation (\ref{grel}), and the $\phi(x)$ will have
the orthonormality property given in Equation (\ref{ortho}). To evaluate the
infinite sum in Equation (\ref{scfuna}), we use the finite Fourier series
expansion of the function
\begin{equation}
g(k) = \sum_{l=-\infty}^\infty \left| \tilde{B}(k+2\pi l)\right|^2 \,. 
             \label{gexp}
\end{equation}
This function has period $2\pi$, and the Fourier expansion has the form
\begin{equation}
g(k) = \sum_{j=-\infty}^\infty c_j e^{ijk} \,. \label{gfour}
\end{equation}
Following the derivation in Equation (\ref{coef}), the expansion
coefficients are given by
\begin{eqnarray}
c_j &=& \frac{1}{2\pi}\int_0^{2\pi} e^{-ijk} g(k)\,dk \nonumber\\
    &=& \frac{1}{2\pi}\int_0^{2\pi} e^{-ijk} \sum_{l=-\infty}^\infty \left| 
            \tilde{B}(k+2\pi l)\right|^2 \,dk \nonumber\\
    &=& \frac{1}{2\pi}\int_{-\infty}^{\infty} e^{-ijk} \left|\tilde{B}(k)
            \right|^2 \,dk \nonumber\\
    &=& \frac{1}{2\pi}\int_{-\infty}^{\infty} \tilde{B}^\ast(k) e^{-ijk}
            \tilde{B}(k) \,dk \nonumber\\
    &=& \int_{-\infty}^{\infty} B^\ast(x) B(x-j) \,dx \,, \label{bcoef}
\end{eqnarray}
where the Parseval relation was used for the last step. The integral 
in Equation
(\ref{bcoef}) is easy to evaluate for the B-splines, and we find
\begin{equation}
\int_{-\infty}^{\infty} B^\ast(x) B(x-j) \,dx 
        = \cases{\frac{1}{120}, &if $j=-2$ \cr
                 \frac{13}{60}, &if $j=-1$ \s\cr
                 \frac{11}{20}, &if $j=0$ \s\cr
                 \frac{13}{60}, &if $j=1$ \s\cr
                 \frac{1}{120}, &if $j=2$ \s\cr
                 0, &otherwise. \s\cr}
\end{equation}
Using these coefficients for the expansion given in Equation (\ref{gfour}), we
find
\begin{equation}
g(k) = \frac{11}{20}+\frac{13}{30}\cos(k)+\frac{1}{60}\cos(2k) \,.
\end{equation}

To find $\phi(x)$ the inverse Fourier transform of $\tilde{\phi}(k)$ must be
done numerically; however, there is a nice method which gives an efficient
algorithm. Since the function $g(k)$ has period $2\pi$, we can use the expansion
\begin{equation}
\frac{1}{\sqrt{g(k)}} = \sum_{n=-\infty}^\infty c_n e^{-ink} \,, \label{rtexp}
\end{equation}
where the coefficients
\begin{equation}
c_n = \frac{1}{2\pi}\int_{-\pi}^{\pi} \frac{e^{ink}}{\sqrt{g(k)}}\,dk 
\end{equation}
must be computed numerically. For the B-spline, the $g(k)$ is an even function
of $k$, and one finds
\begin{equation}
c_n = \frac{1}{2\pi}\int_{-\pi}^{\pi} \frac{\cos(nk)}{\sqrt{g(k)}}\,dk \,.\label{cn}
\end{equation}
In addition, from Equation (\ref{cn}) we see that $c_{-n}=c_n$. Now, using
Equation (\ref{rtexp}) we get
\begin{eqnarray}
\phi(x) &=& \frac{1}{2\pi}\int_{-\infty}^\infty \frac{\tilde{B}(k)}{\sqrt{g(k)}}
             \,e^{ikx}\,dk \nonumber\\
        &=& \frac{1}{2\pi}\int_{-\infty}^\infty \tilde{B}(k) \left[
             \sum_{n=-\infty}^\infty c_n e^{-ink} \right]\,e^{ikx}\,dk
             \nonumber\\
        &=& \sum_{n=-\infty}^\infty c_n \left[\frac{1}{2\pi} 
             \int_{-\infty}^\infty \tilde{B}(k) e^{ik(x-n)}\,dk \right]
             \nonumber\\
        &=& \sum_{n=-\infty}^\infty c_n B(x-n) \,. \label{phi_b}
\end{eqnarray}

Now we need to find the wavelet $\psi(x)$ with the properties
\begin{equation}
\int_{-\infty}^\infty \psi^\ast(x-l) \phi(x-m)\,dx =0 \,,
\end{equation}
and
\begin{equation}
\int_{-\infty}^\infty \psi^\ast(x-l) \psi(x-m)\,dx = \delta_{lm} \,.
\end{equation}
To do this we introduce the functions
\begin{equation}
\phi_{-1,l}(x) = \sqrt{2}\phi(2x-l) \,. \label{phim}
\end{equation}
The Fourier transform of these functions is given by
\begin{eqnarray}
\tilde{\phi}_{-1,l}(k) &=& \int_{-\infty}^\infty e^{-ikx} \phi_{-1,l}(x)\,dx
     \nonumber\\
     &=& \sqrt{2}\int_{-\infty}^\infty e^{-ikx} \phi(2x-l)\,dx \,
\end{eqnarray}
and setting $2x-l=y$ yields
\begin{eqnarray}
\tilde{\phi}_{-1,l}(k) 
    &=& \frac{1}{\sqrt{2}} e^{-ikl/2}\int_{-\infty}^\infty e^{-iky/2} \phi(y)
         \,dy \nonumber\\
     &=& \frac{1}{\sqrt{2}} e^{-ikl/2} \tilde{\phi}(k/2) \,.
\end{eqnarray}

Using the $\phi_{-1,n}(x)$ as an orthonormal basis set, we can write
\begin{equation}
\phi(x) = \sum_{n=-\infty}^\infty h_n \phi_{-1,n}(x) \,, \label{phiexp}
\end{equation}
where
\begin{equation}
h_n = \int_{-\infty}^\infty \phi_{-1,,n}^\ast(x) \phi(x) \,dx \,. \label{hint}.
\end{equation}
This is the scaling equation for this system.  In this case there are an 
infinite number of non-zero scaling coefficients.
Since the $\phi(x-n)$ are orthonormal, the $h_n$ must have the property
\begin{equation}
\sum_{n=-\infty}^\infty \left|h_n\right|^2 = 1 \,.
\end{equation}
Taking the Fourier transform of Equation (\ref{phiexp}) gives
\begin{equation}
\tilde{\phi}(k) = \frac{1}{\sqrt{2}} \sum_{n=-\infty}^\infty h_n e^{-ikn/2}
                    \tilde{\phi}(k/2) \,,
\end{equation}
which can be written as
\begin{equation}
\tilde{\phi}(k) = m_0(k/2)\tilde{\phi}(k/2) \,, \label{prod}
\end{equation}
where
\begin{equation}
m_0(k) = \frac{1}{\sqrt{2}} \sum_{n=-\infty}^\infty h_n e^{-ikn} \,.
\end{equation}
Using Equations (\ref{grel}) and (\ref{prod}) we see that
\begin{eqnarray}
\sum_{l=-\infty}^\infty \left|\tilde{\phi}(2k+2\pi l)\right|^2 
     &=& \sum_{l=-\infty}^\infty \left|\p m_0(k+\pi l)\right|^2 \left|
          \tilde{\phi}(k+\pi l)\right|^2 \nonumber\\     
     &=& \sum_{l,even} \left|\p m_0(k+\pi l)\right|^2 \left|
          \tilde{\phi}(k+\pi l)\right|^2 \nonumber\\
         & & \qquad\qquad {}+ \sum_{l,odd} \left|\p m_0(k+\pi l)\right|^2 \left|
          \tilde{\phi}(k+\pi l)\right|^2 \nonumber\\
     &=& \sum_{l=-\infty}^\infty \left|\p m_0(k+2\pi l)\right|^2 \left|
          \tilde{\phi}(k+2\pi l)\right|^2 \nonumber\\
         & & \qquad\qquad {}+ \sum_{l=-\infty}^\infty \left|\p m_0(k+2\pi l 
         + \pi)\right|^2 \left|\tilde{\phi}(k+2\pi l +\pi)\right|^2 \nonumber\\
     &=& \left|m_0(k)\right|^2 \sum_{l=-\infty}^\infty \left|
          \tilde{\phi}(k+2\pi l)\right|^2 \nonumber\\
         & & \qquad\qquad{}+ \left|m_0(k+\pi)\right|^2 \sum_{l=-\infty}^\infty 
         \left|\tilde{\phi}(k+\pi+2\pi l)\right|^2 \nonumber\\
     &=& \left|m_0(k)\right|^2 + \left|m_0(k+\pi)\right|^2 \label{msum}
\end{eqnarray}
where we have used the periodicity of $m_0(k)$. The sum on the left-hand side
of Equation (\ref{msum}) is equal to unity; thus, we have shown that
\begin{equation}
\left|m_0(k)\right|^2 + \left|m_0(k+\pi)\right|^2 = 1 \label{m0rel}
\end{equation}

Now we use a similar procedure to find $\psi(x)$.
Using the $\phi_{-1,n}(x)$ as an orthonormal basis set, we write
\begin{equation}
\psi(x) = \sum_{n=-\infty}^\infty f_n \phi_{-1,n}(x) \, , \label{psiexpb}
\end{equation}
with
\begin{equation}
f_n = \int_{-\infty}^\infty \phi_{-1,n}^\ast(x) \psi(x) \,dx \,.
\end{equation}
Taking the Fourier transform of Equation (\ref{psiexpb}) gives
\begin{eqnarray}
\tilde{\psi}(k) &=& \frac{1}{\sqrt{2}} \sum_{n=-\infty}^\infty f_n e^{-ikn/2}
                    \tilde{\phi}(k/2) \nonumber\\
                &=& m_1(k/2)\tilde{\phi}(k/2) \,, \label{psiprod}
\end{eqnarray}
where
\begin{equation}
m_1(k) = \frac{1}{\sqrt{2}} \sum_{n=-\infty}^\infty f_n e^{-ikn} \,.
\end{equation}

If $m_1(k)$ has the same property as that given in Equation (\ref{m0rel}) for
$m_0(k)$, then the functions $\psi(x-m)$ will be orthonormal. In addition, we
want $\psi(x-n)$ to be orthogonal to $\phi(x-m)$. Thus, we want to find a
$m_1(k)$ such that
\begin{eqnarray}
\int_{-\infty}^\infty \psi^\ast(x-n) \phi(x-m)
   &=& \frac{1}{2\pi}\int_{-\infty}^\infty e^{ikn}\tilde{\psi}^\ast(k) e^{-ikm}
          \tilde{\phi}(k)\, dk \nonumber\\
   &=& \frac{1}{2\pi}\int_{-\infty}^\infty e^{i(n-m)k}\tilde{\psi}^\ast(k) 
          \tilde{\phi}(k)\, dk \nonumber\\
   &=& \frac{1}{2\pi}\int_0^{2\pi} dk\,e^{i(n-m)k}\sum_{l=-\infty}^\infty 
          \tilde{\psi}^\ast(k+2\pi l) \tilde{\phi}(k+ 2\pi l) \nonumber\\
   &=& 0.
\end{eqnarray}
This condition is satisfied if
\begin{equation}
\sum_{l=-\infty}^\infty \tilde{\psi}^\ast(k+2\pi l) \tilde{\phi}(k+ 2\pi l) 
          = 0 \,. \label{psiorth}
\end{equation}
Substituting Equations (\ref{prod}) and  (\ref{psiprod}) into Equation
(\ref{psiorth}) and replacing $k$ by $2k$ gives
\begin{equation}
\sum_{l=-\infty}^\infty m_1^\ast(k)\tilde{\phi}^\ast(k+\pi l) m_0(k)
     \tilde{\phi}(k+ \pi l) = 0 \,.
\end{equation}
Regrouping the sums for odd and even $l$, and following the procedure used in 
Equation (\ref{msum}) gives
\begin{equation}
m_1^\ast(k) m_0(k) + m_1^\ast(k+\pi) m_0(k+\pi) = 0 \,.
\end{equation}
This condition will be satisfied if we choose
\begin{equation}
m_1(k) = e^{-ik} m_0^\ast(k+\pi) \,.
\end{equation}
Note this choice for $m_1(k)$ is not unique; we can multiply $m_1(k)$ by any
function $\rho(k)$ which has period $\pi$ and $\left|\rho(k)\right|=1$, and
still satisfy the constraints on $m_1(k)$.
Substituting this result into Equation (\ref{psiprod}) gives
\begin{eqnarray}
\tilde{\psi}(k) &=& e^{-ik/2} m_0^\ast(k/2+\pi) \tilde{\phi}(k/2) \nonumber\\
                &=& e^{-ik/2} \frac{1}{\sqrt{2}} \sum_{n=-\infty}^\infty
                       h_n^\ast e^{i(k/2+\pi)n} \tilde{\phi}(k/2) \nonumber\\
                &=& \frac{1}{\sqrt{2}} \sum_{n=-\infty}^\infty (-1)^n h_n^\ast
                      e^{-i(-n+1)k/2} \tilde{\phi}(k/2) \nonumber\\
                &=& \sum_{n=-\infty}^\infty (-1)^n h_n^\ast \tilde{\phi}_{-1,
                     -n+1}(k) \,. \label{psi1}
\end{eqnarray}
Replacing $n$ by $-n+1$ in Equation (\ref{psi1}) gives
\begin{equation}
\tilde{\psi}(k) = - \sum_{n=-\infty}^\infty (-1)^n h_{-n+1}^\ast 
                          \tilde{\phi}_{-1,n}(k) \,.
\end{equation}
For convenience, we drop the minus sign in front of the sum, and write
\begin{equation}
\tilde{\psi}(k) = \sum_{n=-\infty}^\infty g_n \tilde{\phi}_{-1,n}(k) 
                       \,, \label{psi2}
\end{equation}
where $g_n = (-1)^n h_{-n+1}$ for $h_n$ a real number.
Taking the Fourier transform of Equation (\ref{psi2}) gives the result
\begin{eqnarray}
\psi(x) &=& \sum_{n=-\infty}^\infty g_n \phi_{-1,n}(x) \nonumber\\
        &=& \sqrt{2} \sum_{n=-\infty}^\infty g_n \phi(2x-n) \,. \label{pseries}
\end{eqnarray}

To evaluate $\psi(x)$ we use the expansion given in Equation (\ref{phi_b}) in
Equation (\ref{pseries}). This gives
\begin{equation}
\psi(x) = \sqrt{2} \sum_{n=-\infty}^\infty g_n \sum_{m=-\infty}^\infty c_m
                 B(2x-n-m) \,.
\end{equation}
Now replace $m$ by $l-n$, this gives
\begin{eqnarray}
\psi(x) &=& \sum_{l=-\infty}^\infty \left[\sqrt{2} \sum_{n=-\infty}^\infty g_n
                 c_{l-n} \right] B(2x-l) \nonumber\\
        &=& \sum_{l=-\infty}^\infty d_l B(2x-l) \,.
\end{eqnarray}

\centerline{\sc Numerical Methods}
\medskip
To determine the $h_n$ we need to evaluate the overlap integral given in
Equation (\ref{hint}). From Equations (\ref{phim}) and (\ref{phi_b}), we find
\begin{equation}
\phi_{-1,n}(x) = \sqrt{2}\sum_{m=-\infty}^\infty c_m B(2x-m-n) \,.
\end{equation}
Then using 
\begin{equation}
B(x) = \sum_{j=-\infty}^\infty b_j B(2x-j) \,, \label{bscale}
\end{equation}
where, the $b_j$ for the quadratic B-splines are given by
\begin{equation}
b_j = \cases{\frac{1}{4}, &if $j=-1$ \cr
             \frac{3}{4}, &if $j=0$ \s\cr
             \frac{3}{4}, &if $j=1$ \s\cr
             \frac{1}{4}, &if $j=2$ \s\cr
                0, &otherwise. \s\cr} \,,
\end{equation}
we can write
\begin{eqnarray}
\phi(x) &=& \sum_{n=-\infty}^\infty c_n \sum_{j=-\infty}^\infty b_j B(2x-2n-j)
             \nonumber\\
        &=& \sum_{l=-\infty}^\infty \left[ \sum_{n=-\infty}^\infty c_n b_{l-2n}
             \right] B(2x-l) \nonumber\\
        &=& \sum_{l=-\infty}^\infty s_l B(2x-l) \,.
\end{eqnarray}
Using these expansions, the overlap integral is given by
\begin{eqnarray}
h_n &=& \sqrt{2}\sum_{m=-\infty}^\infty \sum_{l=-\infty}^\infty c_m s_l
        \int_{-\infty}^\infty B(2x-m-n) B(2x-l) \,dx \nonumber\\
    &=& \sqrt{2}\sum_{m=-\infty}^\infty \sum_{l=-\infty}^\infty c_m s_l
        \frac{1}{2}\int_{-\infty}^\infty B(x-m-n+l) B(x) \,dx \nonumber\\
    &=& \frac{1}{\sqrt{2}}\,\sum_{m=-\infty}^\infty \sum_{j=-2}^2 c_m s_{m+n-j}
        \int_{-\infty}^\infty B(x) B(x-j) \,dx \,,
\end{eqnarray}
where, we have set $l=m+n-j$ in the second summation. The values for the integrals
of the quadratic B-splines are given in Equation (\ref{bcoef}).

To derive Equation (\ref{bscale}), we use $\sin(2\theta)=2 \cos(\theta)
\sin(\theta)$ to write Equation (\ref{qbk}) as
\begin{eqnarray}
\tilde{B}(k) &=& e^{-ik/2} \left[\frac{\cos(k/4)\sin(k/4)}{k/4}\right]^3
                   \nonumber\\
             &=& e^{-ik/4}\left(\frac{e^{ik/4}+e^{-ik/4}}{2}\right)^3 e^{-ik/2}
                  \left[\frac{\sin(k/4)}{k/4}\right]^3 \nonumber\\
             &=& e^{-ik/4}\left(\frac{e^{ik/4}+e^{-ik/4}}{2}\right)^3 
                  \tilde{B}(k/2) \nonumber\\
             &=& \left(\frac{e^{ik/2} + 3 + 3\,e^{-ik/2} + e^{-ik}}{8}\right)
                    \tilde{B}(k/2) \,. \label{b3exp}
\end{eqnarray}
Now taking the Fourier transform of (\ref{b3exp}) and using
\begin{eqnarray}
\int_{-\infty}^\infty e^{-ikx} B(2x-j)\,dx &=& \frac{1}{2}\,e^{ikj/2} 
       \int_{-\infty}^\infty e^{-ikx/2} B(x)\,dx \nonumber\\
      &=& \frac{1}{2}\,e^{ikj/2}B(k/2) \,,
\end{eqnarray}
we find
\begin{equation}
B(x) = \frac{1}{4} B(2x-1) + \frac{3}{4}B(2x) + \frac{3}{4}B(2x+1)
         + \frac{1}{4} B(2x-2) \,.
\end{equation}

\bigskip
\hrule 
\bigskip 
The authors would like to thank Mr. Fatih Bulut for a careful
proof reading of these notes. 

\end{document}